\documentclass{pasj01}
\Received{$\langle$reception date$\rangle$}
\Accepted{$\langle$acception date$\rangle$}
\Published{$\langle$publication date$\rangle$}
\usepackage{color}

\usepackage{multirow}
\usepackage{url}

\usepackage[switch,mathlines]{lineno}

\begin{document}

\title{Gravitational Collapse of White Dwarfs to Neutron Stars.\\
From Initial Conditions to Explosions with Neutrino-radiation Hydrodynamics Simulations}

\author{Masamitsu Mori\thanks{These two authors contributed equally to this work.}\altaffilmark{1,2}}
\altaffiltext{1}{Department of Earth Science and Astronomy, Graduate School of Arts and Sciences, The University of Tokyo, Tokyo 153-8902, Japan}
\altaffiltext{2}{National Astronomical Observatory of Japan, 2-21-1, Osawa, Mitaka, Tokyo, 181-8588, Japan}
\email{masamitsu.mori.astro@gmail.com}

\author{Ryo Sawada\thanks{These two authors contributed equally to this work.}\altaffilmark{1}}
\email{ryo@g.ecc.u-tokyo.ac.jp}

\author{Yudai Suwa\altaffilmark{1,3}}
\altaffiltext{3}{Center for Gravitational Physics and Quantum Information, Yukawa Institute for Theoretical Physics, Kyoto University, Kyoto 606-8502, Japan}

\author{Ataru Tanikawa\altaffilmark{1,4}}
\altaffiltext{4}{Center for Information Science, Fukui Prefectural University, 4-1-1 Matsuoka Kenjojima, Eiheiji-cho, Fukui, 910-1195, Japan}

\author{Kazumi Kashiyama\altaffilmark{5,6}}
\altaffiltext{5}{Astronomical Institute, Graduate School of Science, Tohoku University, Aoba, Sendai 980-8578, Japan}
\altaffiltext{6}{Kavli Institute for the Physics and Mathematics of the Universe (Kavli IPMU,WPI), The University of Tokyo, Chiba 277-8582, Japan}

\author{Kohta Murase\altaffilmark{3,7,8}}
\altaffiltext{7}{Department of Physics; Department of Astronomy \& Astrophysics; Center for Multimessenger Astrophysics, Institute for Gravitation and the Cosmos, The Pennsylvania State University, University Park, PA 16802, USA}
\altaffiltext{8}{School of Natural Sciences, Institute for Advanced Study, Princeton, NJ 08540, USA}


\maketitle

\begin{abstract}
This paper provides collapses of massive, fully convective, and non-rotating white dwarfs (WDs) formed by accretion-induced collapse or merger-induced collapse and the subsequent explosions with the general relativistic neutrino-radiation hydrodynamics simulations with the M1 multi-energy scheme in one-dimension. We produce initial WDs in hydrostatic equilibrium, which have super-Chandrasekhar mass and are about to collapse. The WDs have masses of 1.6$M_\odot$ with different initial central densities specifically at $1.0\times10^{10}$, $4.0\times10^{9}$, $2.0\times10^{9}$ and $1.0\times10^{9}\,{\rm g\,cm^{-3}}$. First, we check whether initial WDs are stable without weak interactions. Second, we calculate the collapse of WDs with weak interactions. We employ hydrodynamics simulations with Newtonian gravity in the first and second steps. Third, we calculate the formation of neutron stars and accompanying explosions with general relativistic simulations. As a result, WDs with the highest density of $10^{10}\,{\rm g\,cm^{-3}}$ collapse not by weak interactions but by the photodissociation of the iron, and three WDs with low central densities collapse by the electron capture as expected at the second step and succeed in the explosion with a small explosion energy of $\sim 10^{48}$ erg at the third step. 
By changing the surrounding environment of WDs, we find that there is a minimum value of ejecta masses being $\sim 10^{-5}M_{\odot}$. With the most elaborate simulations of this kind so far, the value is one to two orders of magnitude smaller than previously reported values and is compatible with the estimated ejecta mass from FRB~121102.
\end{abstract}


\section{Introduction}\label{sec:intro}
Accretion-induced collapse (AIC) is a theoretically predicted phenomenon in which a white dwarf (WD) collapses into a neutron star (NS). 
This collapse can trigger a variety of explosive transients, which may have been observed throughout the universe. 
Revealing the AIC may give us insights into the evolution of binary star systems, the behavior of dense matter, and the physics of extreme environments. 
In this context, understanding the mechanics and outcomes of AIC is essential for gaining a more comprehensive understanding of the universe and its diverse structures. 

There are two pathways to the formation of super-Chandrasekhar WDs.
The first is the AIC, and the second is the merger-induced collapse (MIC). 
The difference between the AIC and the MIC is whether single degenerate or double degenerate form super-Chandrasekhar WDs. 
The former is the AIC, and the latter is the MIC. 
Whichever path forms NSs, the NSs, however, evolve in the same way. 
Hence we use the term `AIC' to indicate both means `AIC' and `MIC'.

A promising progenitor of AIC is a merger remnant of binary carbon-oxygen WDs (e.g., \cite{2020IAUS..357....1R}). Such a merger remnant may cause a type Ia supernova if it slowly accretes from a tidally disrupted WD \citep{2007MNRAS.380..933Y}. However, magnetohydrodynamic simulations have reported that magnetic viscosity leads to a high accretion rate and ignites carbon deflagration \citep{2012MNRAS.427..190S, 2013ApJ...773..136J}.
The carbon deflagration entirely converts materials of a merger remnant from carbon-oxygen to oxygen-magnesium-neon or heavier elements \citep{1985A&A...150L..21S, 2016MNRAS.463.3461S}. If the merger remnant has super-Chandrasekhar mass, it will cause AIC \citep{1991ApJ...367L..19N}.
The above scenario that a WD merger evolves into a super-Chandrasekhar oxygen-magnesium-neon WD has been theoretically pointed out for a long time. Only recently, a candidate of the scenario has been discovered \citep{2019Natur.569..684G, 2019ApJ...887...39K, 2020A&A...644L...8O, 2023ApJ...944..120L, 2023arXiv230414669K}.

AIC may have connections to unresolved astrophysical phenomena, such as fast radio bursts (FRBs; \cite{2017ApJ...839L...3K,2022Natur.602..585K}) and the Galactic Center GeV Excess (GCE; \cite{2022NatAs...6..703G}).
Historically, AIC has been proposed to explain a variety of troublesome NS-forming systems, such as millisecond pulsars in globular clusters and galactic disks \citep{1990ApJ...353..159B,1990ARA&A..28..183C}. 
And the Fast Blue Optical Transients (FBOTs), which were discovered in recent years but are difficult to explain by supernovae (e.g., \cite{2022ApJ...927..223S}), have also been suggested in relation to AIC \citep{2021ApJ...922..247O,2023MNRAS.518..623M}.
A better understanding of the physical processes of AICs will provide a deeper understanding of the universe's various mysterious and energetic issues.

As used in this context, the term `AIC simulation' denotes the calculation from core collapse triggered by electron capture to the explosion.
The pioneering AIC simulation is \cite{1987ApJ...320..304B}, followed by \cite{1992ApJ...391..228W} and \cite{1999ApJ...516..892F}.
While the treatment of microphysics is simple, the emphasis on multidimensional hydrodynamic simulation was led by \cite{2006ApJ...644.1063D}, \cite{2010PhRvD..81d4012A}, and more recently by \cite{2023MNRAS.525.6359L}.
And in the 2020s, 1D simulations have been reported that treat microphysics more seriously than before, although the neutrino transport scheme remains highly approximated. For exmaple, \cite{2020ApJ...894..146S} did not solve neutrino transport but calculated deleptonization with the analytic scheme. \cite{2023MNRAS.518..623M} employed the diffusion approximation.
All these previous studies reported that WDs get unstable by electron capture, collapse, and explode.

On the other hand, this work has the following three notable features from previous studies:
First, our simulations start from exactly equilibrium initial models to reproduce quasi-statistic evolution and it ensures that the WD collapse is truly caused by electron capture.
The stability of the initial WD model has been excessively neglected in all previous studies, while our conclusion is based on its careful evaluation, through which we avoid non-physical gravitational instabilities of massive WDs.
Note that our calculation starts from nuclear statistical cores, which can no longer cause type Ia supernovae.
Second, some previous studies partially omitted neutrino physics~\citep{2006ApJ...644.1063D, 2010PhRvD..81d4012A}. We incorporate a more accurate neutrino transport scheme, in which we calculate neutrinos of multi flavors and multi-energy groups in the general relativistic gravity. We also calculate both neutrino-neutrino and neutrino-matter interaction considering microphysics. Our scheme is more accurate than that of \cite{2023MNRAS.525.6359L}. See Section~\ref{subsec:neutrino_trans} for details.
\cite{2020ApJ...894..146S} has suggested that neutrino handling changes the dynamics of the AIC, contributing to the accuracy and reliability of our results.
Third, except for a few examples \citep{2023MNRAS.525.6359L,2010PhRvD..81d4012A}, the past works have employed Newtonian gravity, whereas this study takes into account general relativity (GR). The role of GR cannot be overlooked, especially when dealing with the NS formation.

We simply describe the pathway from two WD mergers to a super-Chandrasekahr WD in hydrostatic. See Section~\ref{subsec:wd-bg} for the details of the physical background of our initial models. A binary with two WDs gradually lose its orbital energy and angular momentum through gravitational waves. The lighter WD is disrupted by the tidal force and the heavier WD is surrounded by a Keplerian disk. The Keplerian disk gradually accretes onto the heavier WD. If silicon-group elements are formed, the WD can occur core-collapse instead of a type Ia supernova \citep{2015A&A...580A.118M}.

In this paper, we perform a self-consistent simulation of a WD from its hydrostatic structure, through its collapse by electron capture reactions, to the explosion and the formation of proto-neutron stars (PNSs). We describe our initial model in Section~\ref{sec:model}, our simulation setup in Section~\ref{sec:hydro}. In order to check the stability of initial models, we carry out simulation without weak interactions in Section~\ref{subsec:adiabatic}. Our simulation results are shown in Section~\ref{sec:results}. Finally, we provide a summary and discussion in Section~\ref{sec:discus}.
we use the natural units: $G=c=1$ and the Minkoswki metric as the set of signs: $(-,+,+,+)$. 

\section{Initial model}\label{sec:model}
In this study, we do not evolve a binary system and we assume, as initial conditions of super-Chandrasekhar WDs, with a total mass of about $1.6 M_\odot$ which has achieved hydrostatic equilibrium with central densities of $1.0\times10^{9}$\,g\,cm$^{-3}$, $2.0\times10^{9}{\rm \,g\,cm^{-3}}$, $4.0\times10^{9}{\rm \,g\,cm^{-3}}$ and  $1.0\times10^{10}{\rm \,g\,cm^{-3}}$ and the interiors of WDs to be fully convective. They are summarized in Table~\ref{tab:initial_models}. 
The mass of $1.6 M_\odot$ is heavy for normal cold WDs partially because there is a difficulty of simulations of light WDs (See Section~\ref{sec:summary}). However, we deal with hot and heavy WDs right after accretion in the AIC. Hence, this assumption is reasonable.

While we consider these models to be remnants produced by the two WD mergers, we do not explicitly discuss the physical trajectory at which this initial condition is achieved in this paper.
Here, to avoid any confusion, we recapitulate what evolutionary pathway and which phases of the super-Chandrasekhar WD are assumed as models in this study in Section~\ref{subsec:wd-bg}, and we then summarize the physical profiles of the super-Chandrasekhar WDs used as initial conditions in Section~\ref{subsec:wd-model}.

\subsection{Background on Initial Model}\label{subsec:wd-bg}
This subsection details the pathway from a two WDs merger to a super-Chandrasekhar WD in hydrostatic equilibrium. 
Here, we first assume two WDs to be two carbon-oxygen WDs with a total mass of $1.6M_\odot$. A binary with two WDs gradually lose its orbital energy and angular momentum through gravitational wave radiation. The binary orbit is finally circularized when the lighter WD fills its Roche lobe and starts Roche lobe overflow. The Roche lobe overflow is unstable if the mass ratio of the lighter WD to the heavier WD is large, say $\gtrsim 0.2$ (e.g., \citep{2004MNRAS.350..113M}). It eventually leads to tidal disruption of the lighter WD.\footnote{Type Ia supernovae or their variants can occur before, during, and shortly after the tidal disruption according to the helium-ignited violent merger or dynamically driven double degenerate double detonation models \citep{2010ApJ...709L..64G, 2013ApJ...770L...8P, 2021MNRAS.503.4734P, 2018ApJ...865...15S, 2018ApJ...868...90T, 2019ApJ...885..103T}, carbon-ignited violent merger model \citep{2010Natur.463...61P, 2011A&A...528A.117P, 2012MNRAS.424.2222P, 2012ApJ...747L..10P, 2015ApJ...807...40T}, and spiral instability model \citep{2015ApJ...800L...7K, 2018ApJ...869..140K}, respectively. Calcium-rich supernovae may be also possible \citep{2010Natur.465..322P, 2019arXiv191007532P, 2022arXiv220713110Z}.} 

Then, the merger remnant has a cold core surrounded by a hot envelope and Keplerian disk, where the cold core is made of the heavier WD, and the hot envelope and Keplerian disk are debris of the lighter WD \citep{1990ApJ...348..647B, 2007MNRAS.380..933Y, 2011ApJ...737...89D, 2014ApJ...788...75R}. Due to magnetic viscosity, the Keplerian disk accretes onto the cold core and hot envelope. This accretion process ignites carbon deflagration in the hot envelope \citep{2012MNRAS.427..190S, 2013ApJ...773..136J}.\footnote{Dynamical accretion process following the tidal disruption can ignite carbon deflagration \citep{2015ApJ...807..105S, 2016ApJ...821...67S}.} The carbon deflagration invades the cold core and converts carbon-oxygen materials into oxygen-neon-magnesium materials \citep{1985A&A...150L..21S}.
Viscous heating between the core and the envelope also provides thermal energy and makes the temperature profile convective \citep{2012MNRAS.427..190S}. 
Then, neon ignition occurs, and silicon-group elements are formed \citep{2016MNRAS.463.3461S}.
Finally, at some point, it gravitationally collapses to a NS through electron capture onto silicon-group elements \citep{1991ApJ...367L..19N}. 

In this way, we assume the formation of super-Chandrasekhar WDs, which are convection-dominated and composed of heavy elemental material, as the initial conditions.
Note that such a WD should not cause type Ia supernovae; Although \cite{2015A&A...580A.118M} have shown type Ia supernovae from WDs consisting of oxygen-neon-magnesium materials, they have considered sub-Chandrasekhar WDs, which are stable against the gravitational collapse.

\subsection{Setup of Initial Model}\label{subsec:wd-model}

The initial conditions for hydrodynamic simulations are obtained by hydrostatic equilibrium calculations of a hot WD with an adiabatic temperature gradient based on open code~\citep{adiabatic_wd}. The mass of our initial model is $1.6M_\odot$. The mass is heavier than the standard WD. The reasons why we emply the mass are first that we consider a hot WD after accretion and second that simulation of core collapse of light WDs is time-consuming (see Section~\ref{sec:discus}).
We integrate the relevant equations of stellar structure as 
\begin{eqnarray}
    \frac{d r}{dm} &=& \frac{3}{4\pi r^2\rho} \label{eq:wd1}~,\\
    \frac{d P}{dm} &=& -\frac{Gm\rho}{r^2}\frac{d r}{dm} \label{eq:wd2}~,\\
    \frac{d T}{dm} &=& \frac{d r}{dm} \frac{T}{P} \left( \frac{\partial \mathrm{ln}T}{\partial \mathrm{ln}P} \right)_\mathrm{ad} \label{eq:wd3}~,
\end{eqnarray}
where $\rho$, $T$, $P$, $r$, and $m$ are density, temperature, total pressure, radius, and enclosed mass, respectively and the subscript of ``ad'' means ``adiabatic''.
To determine temperature, the code assumes fully convective WDs, which leads to Eq.~\ref{eq:wd3}. According to \cite{2016MNRAS.463.3461S, 2021ApJ...906...53S}, the entropy gradient of the merged WD is small. In particular, the entropy profile is constant in the envelope.

These equations are closed by the Timmes equation of state \citep{2000ApJS..126..501T} 
and neglect the effects of nuclear burning on hydrostatic equilibrium. Not considering them does not affects dynamics of WDs because the contribution of electron gas is dominant.
The electron and positron gas is more dominant to pressure than the number of baryon at the density of $10^9{\rm \,g\,cm^{-1}}$. Nuclear burning also releases nuclear energy. However, the nuclear energy is $3.2\times10^{50}$\,erg~\citep{2018ApJ...869..140K} and is more than 10 times smaller than that of binding energy, which is $10^{52}$\,erg for our models. Therefore, the influence of nuclear burning on the dynamics of collapse is negligible.

\begin{figure*}
\centering
  \includegraphics[width=0.8\textwidth]{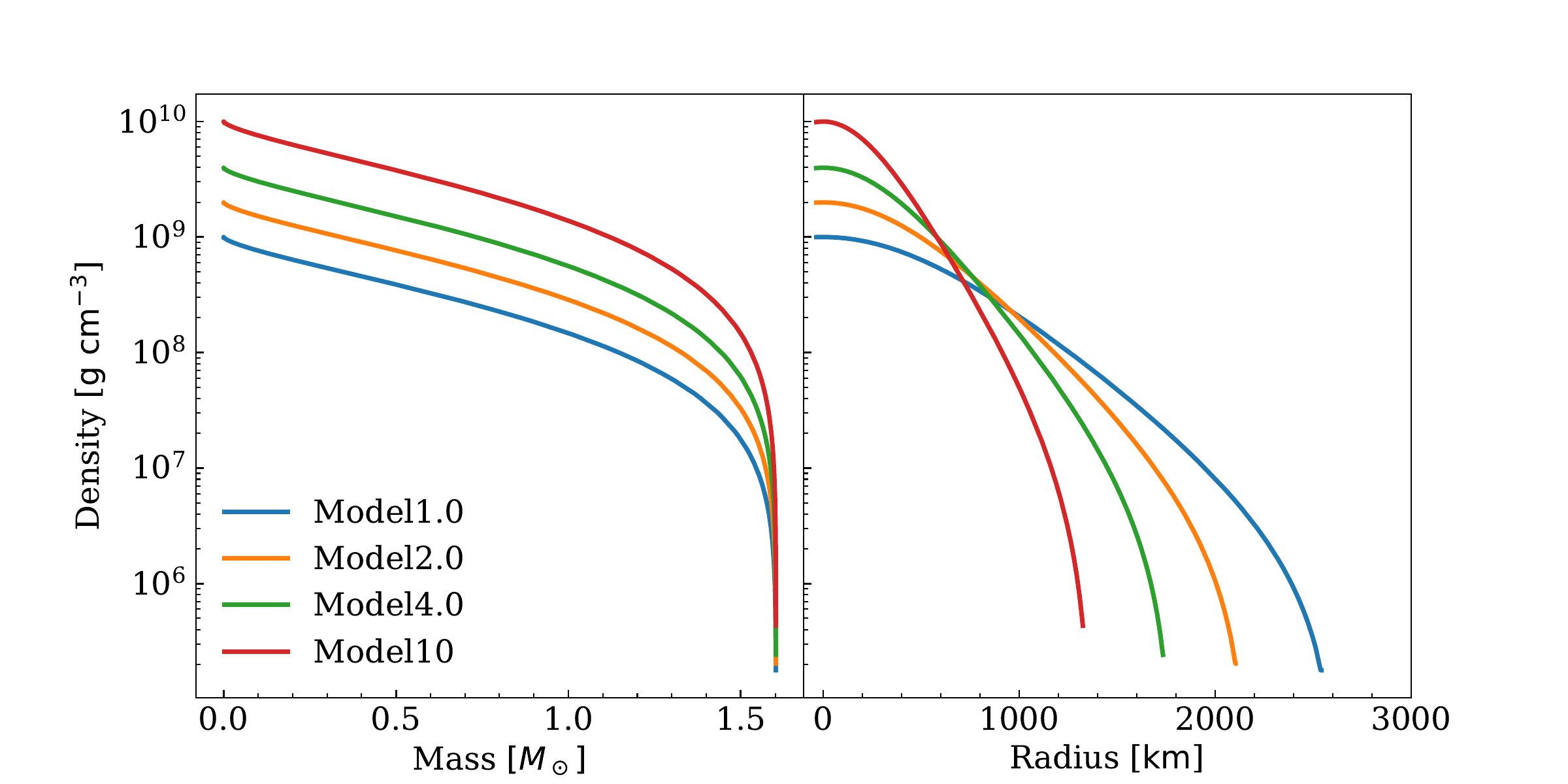}
\caption{Density structure as a function of the enclosed mass and radius for initial models of super-Chandrasekhar WDs.
Each color corresponds to WD models Model1.0 (blue), Model2.0 (orange), Model4.0 (green) and Model10 (red).
The color convention for each model is kept the same throughout our paper.}
\label{fig:WD}
\end{figure*}

Figure~\ref{fig:WD} shows the result of the density structure of the super-Chandrasekhar WDs as a function of the enclosed mass.
Table~\ref{tab:initial_models} summarizes the properties of the intial models.
\begin{table}
    \centering
    \caption{Initial models} 
    \begin{tabular}{cccc}
    \hline\hline
    Mass & Model name &Initial $\rho_c$ & Maximum radius \\
   \lbrack $M_\odot$\rbrack & & \lbrack cm$^{-3}$\rbrack  & \lbrack km\rbrack \\ \hline
    \multirow{4}{*}{1.6} & Model10 &$1.0\times10^{10}$ & $1300$ \\
    & Model4.0 &$4.0\times10^{9}$ & $1700$ \\
     & Model2.0& $2.0\times10^{9}$ & $2100$\\
     &Model1.0& $1.0\times10^{9}$ & $2500$  \\
    \hline\hline
    \end{tabular}
    \label{tab:initial_models}
\end{table}

\begin{figure}
    \centering
    \includegraphics[width=0.49\textwidth]{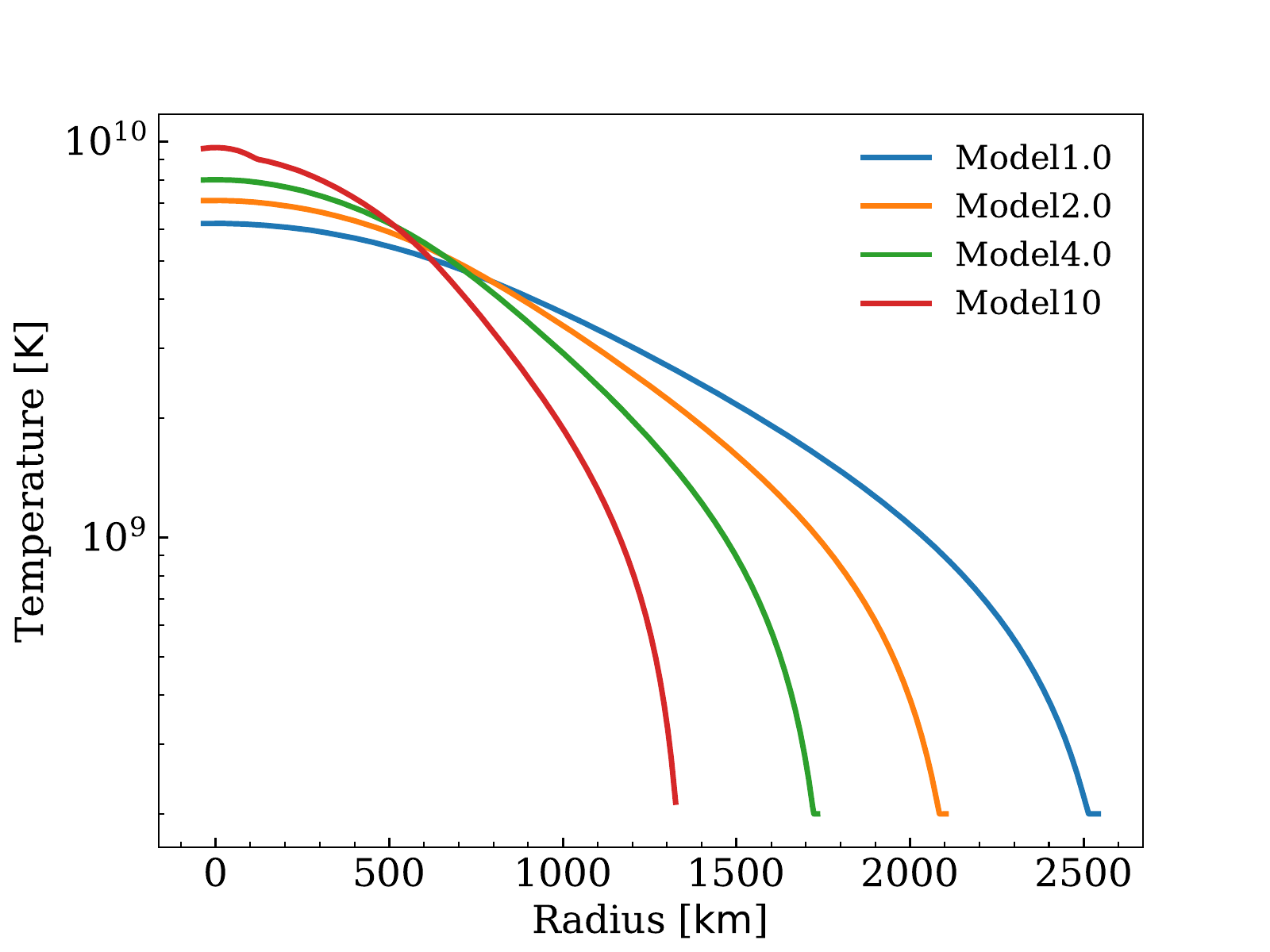}
    \caption{Temperature profiles of the initial WD models. The vertical axis is the radius.}
    \label{fig:temp_profiles}
\end{figure}

Figure~\ref{fig:temp_profiles} shows the initial temperature profiles of WDs. The temperature is above $5\times10^{9}$\,K at the center and decreases to $10^{8}$\,K at the outer region.
The entropy profiles of WDs are around 1\,$k_{\rm B}$ per baryon, which is similar to ordinary iron cores.
Since the equation of state employed in hydrodynamics simulation assumes that the baryons are composed of Ni due to the nuclear statistical equilibrium, the hydrostatic WDs are also assumed to be composed of Ni. This assumption does not change the structure much because the baryon contribution to the total pressure is small. For instance, the electron pressure is $P_e= 4.9\times 10^{26}(\rho /10^{9}{\rm g\,cm^{-3}})^{4/3}(Y_e/0.5)^{4/3}$ dyn\,cm$^{-2}$, where $Y_e$ is the electron fraction, while the baryon pressure is $P_b= 1.5\times 10^{24}(\rho/10^9{\rm g\,cm^{-3}})(T/10^{10}{\rm K})(A/56)^{-1}$\,dyn\,cm$^{-2}$, where $A$ is the mass number and $Z$ is the atomic number (see Section 2 of \cite{1983bhwd.book.....S}).

The pre-collapse configuration depends on the cooling process of the merger remnant after the carbon deflagration. Since the neutrino cooling timescale is shorter in a more dense region \citep{1996ApJS..102..411I}, the center material cools faster, and it forms a WD in hydrostatic equilibrium at the center of the merger remnant. And then, the WD gradually grows with the surrounding materials cooling and accreting onto the WD. Just before the collapse, the merger remnant may consist of a near-Chandrasekhar ($\sim 1.4M_\odot$) WD in hydrostatic equilibrium and its surrounding hot envelope with $\sim 0.2M_\odot$, not a super-Chandrasekhar (here $1.6M_\odot$) WD in hydrostatic equilibrium. Nevertheless, the situation might not be simple like that. Neutrino cooling might proceed simultaneously outside of the WD to some extent, depending on the temperature structure of the merger remnant after the carbon deflagration. In that case, a super-Chandrasekhar WD in hydrostatic equilibrium might be archived.

The feasibility of a super-Chandrasekhar WD is unclear. However, our numerical simulation has one advantageous point over previous simulations (e.g., \cite{1991ApJ...367L..19N}). Our numerical simulation follows WD collapse driven by the electron capture in NSE material. 
We initially select initial models, which is stable without electron capture and unstable with electron capture, in order to simulate collapse triggered by electron capture in section~\ref{subsec:adiabatic}. The procedure allows us to rule out initial models which collapse triggered by the other process, for example, photodissociation in section~\ref{subsec:collapse}.
A super-Chandrasekhar WD with high density does not take time to start its collapse. Note that there is an exception, which employed a lower density model of $4\times10^{9}\,$g\,cm$^{-3}$. It took 200\,ms for the low density model to collapse. We check stability of WD for longer time~\cite{2010PhRvD..81d4012A}. Our numerical simulation should reproduce the onset of WD collapse more correctly than previous simulations.

\section{Simulation Setup}\label{sec:hydro}
\subsection{Simulation Overview}
Our simulation uses the open source code for 1D core-collapse, {\tt GR1D} \citep{2010CQGra..27k4103O, 2015ApJS..219...24O}. {\tt GR1D} is implemented with both the Newtonian and the GR gravity and includes the M1 scheme for the neutrino-radiation transport, which have been used in \cite{2021PTEP.2021b3E01M,2023PhRvD.107h3015M} to conduct simulations for core-collapse supernovae from iron cores.

Our simulations are performed in three steps: 
First, we check the validity of the hydrostatic equilibrium for initial conditions using Newtonian hydrodynamic calculations without the weak interactions (Section~\ref{subsec:adiabatic}).
Then, by taking into account weak interactions, we calculate the process from triggering gravitational instability to the collapse of the WD (Section~\ref{subsec:collapse}, `collapse phase'), 
and finally, we calculate the explosion process using GR hydrodynamics simulations with neutrino transport (Section~\ref{subsec:explosion}, `explosion phase').

The computational domain is taken from 0 km to the radius at which the density structure of the initial WD decreases to $10^5$\,g\,cm$^{-3}$. The maximum radius is between $1.5-2.0\times 10^8$ cm.
In our study, we perform Newtonian simulations on a grid comprising six uniformly spaced zones, extending up to 60 km, and an additional 594 radial zones that are logarithmically spaced, reaching up to the radius where density reaches $10^5{\,\rm g\,cm^{-3}}$. The maximum radius is about 2,000 km. Conversely, the GR simulations are carried out on a grid with 40 uniformly spaced zones spanning up to 20 km and logarithmically spaced radial zones in the outer region. The maximum radius in the GR simulations is 10,000 km, and we have to extend 
the WD profiles to it. Hence, we put the low-density atmosphere of $2.0{\,\rm g cm^{-3}}$. See Section~\ref{sub:EoS} for a detailed discussion of the influences of the atmosphere.

Of particular note is the treatment of the transition from calculations in Newtonian to GR. When the central density reaches  $10^{10}$\,g\,cm$^{-3}$ while calculating the collapse phase, we captured snapshots of the density, velocity, temperature, and electron fraction as a function of radius. We then reconstruct new initial conditions and start simulations with GR.

\subsection{Metric}\label{sec:gr1d}
The metric of {\tt GR1D} is the following:
\begin{equation}
    ds^2 = -\alpha(r, t)^2 dt^2 + X(r,t)^2 dr^2 + r^2d\Omega^2, 
\end{equation}
where $\alpha$ and $X$ are a lapse and a shift function, and we need functions of a potential $\Phi$ and an enclosed gravitational mass $m(r,t)$ to decide them as,
\begin{eqnarray}
    \alpha &=& {\rm exp}(\Phi(r, t)), \\
    X &=& \sqrt{1 - \frac{2m(r,t)}{r}}.
\end{eqnarray}
Here we define enthalpy $h=1+\epsilon+P/\rho$ and Lorentz factor $W=1/\sqrt{1-v}$, where $P$ is pressure, $v$ is velocity $v\equiv X\frac{\partial r}{\partial t}$ and $\epsilon$ is specific internal energy. 
Then, the enclosed gravitational mass and the potential read
\begin{eqnarray}
    m(r,t) =& 4\pi\int^r_0 (\rho h W^2 - P + \tau^\nu_m)r^{\prime 2} dr^\prime,\\
    \Phi(r,t) = &\Phi_0 + \\ \nonumber
    &\int^r_0X^2\left[\frac{m(r^\prime,t)}{r^{\prime 2}} +  4\pi r^\prime(\rho h W^2 v^2 + P + \tau^\nu_\Phi)\right]dr^{\prime 2},
\end{eqnarray}
where $\tau^\nu_m$, $\tau^\nu_\Phi$ are due to the energy and pressure of neutrinos and $\Phi_0$ is determined by the matching condition.
The metric must be connected to the Schwarztschild metric at the surface of the star, which leads to
\begin{equation}
    \Phi(R_*, t) = {\rm ln}[\alpha(R_*, t)] = \frac{1}{2}{\rm ln}\left[1 - \frac{2m(R_*, t)}{R_*}\right],
\end{equation}
where $R_*$ is the radius of the star.

\subsection{Hydrodynamics equations}
The hydrodynamics equations implemented in {\tt GR1D} are abstracted as below,
\begin{eqnarray}
    \partial_t \vec{U} + \frac{1}{r^2}\partial_r\left[\frac{\alpha r^2}{X}\vec{F}\right] = \vec{S},
\end{eqnarray}
where $\vec{U}$ is a vector of conserved values, $\vec{F}$ is a vector of flow values, $\vec{S}$ is a vector of source terms and $\partial_x$ is the same as $\frac{\partial}{\partial x}$.
There are four hydrodynamics equations in {\tt GR1D}.
The first equation is a continuity equation, the second equation is the conservation of leptons, the third equation is the conservation of momentum and the last equation is the conservation of energy.
The compositions of $\vec{U}$ are also abstracted
\begin{eqnarray}
    \vec{U}=[D, DY_e, S^r, \tau],
\end{eqnarray}
The flux vector $\vec{F}$ is
\begin{eqnarray}
    \vec{F}=[Dv, DY_e v, S^rv + P, S^r -Dv],
\end{eqnarray}
In {\tt GR1D}, changing compositions of these abstract vectors allows us to switch from Newtonian gravity to GR.
In GR, the compositions of $\vec{U}$ read
\begin{eqnarray}
    D &=& X\rho W,\\
    DY_e &=& X\rho W Y_e,\\
    S^r &=& \rho h W^2 v, \\
    \tau &=& \rho h W^2 - P - D.
\end{eqnarray}
The source and sink terms read
\begin{eqnarray}
    \vec{S}=\left[0,~R^\nu_{Y_e},~(S^rv-\tau-D)\alpha X\left(8\pi r P + \frac{m}{r^2}\right)\right.\\\nonumber
    \left. + \alpha P X\frac{m}{r^2} + \frac{2\alpha P}{X r} + Q^{\nu, E}_{S^r} + Q^{\nu, M}_{S^r},\right.\\\nonumber 
    \left. Q^{\nu, E}_{\tau} + Q^{\nu, M}_{\tau}\right],
\end{eqnarray}
where $R^\nu_{Y_e}$ is a source term of leptonic number, $Q^{\nu, E}_{S^r}$ and $Q^{\nu, M}_{S^r}$ are source terms of momentum, and $Q^{\nu, E}_{\tau}$ and $Q^{\nu, M}_{\tau}$ are source terms of energy, respectively. The reason why there are two source terms of $Q^{\nu, M}$ and $Q^{\nu, E}$ is because we have to move the fluid frame and the lab frame during neutrino radiation transport, so the energy and the momentum are mixed.
In the Newtonian limit, we assume three conditions: gravity is weak enough, velocity is slow enough, and pressure and specific internal energy are small enough.
That is, when we express the assumption mathematically, the first condition leads to 
\begin{eqnarray}
    g^{\mu \nu} &=& \eta^{\mu \nu} + h^{\mu\nu},\label{eq:metric_weak} \\
    |h^{\mu\nu}| &\ll& 1,\label{eq:weak_enough}
\end{eqnarray}
where $\eta^{\mu\nu}$ is the Minkowski metric and $h^{\mu\nu}$ is  small perturbation from the flat spacetime,
the second condition demands
\begin{eqnarray}
    v_1 \equiv \frac{\partial r}{\partial t} \ll  1,
\end{eqnarray}
and the last condition is
\begin{eqnarray}
        P/\rho \ll 1, \\
    \epsilon \ll 1.
\end{eqnarray}
We ignore higher orders of small terms and then get compositions of $\vec{U}$ in the Newtonian gravity
\begin{eqnarray}
    D &=& \rho,\\
    DY_e &=& \rho Y_e, \\
    S^r &=& \rho v_1, \\
    \tau &=& \rho \epsilon + \frac{1}{2}\rho v_1^2.
\end{eqnarray}
As the flux vector $\vec{F}$, we get
\begin{eqnarray}
    \vec{F}=[Dv_1, DY_e v_1, S^rv_1 + P, (\tau + P)v_1],
\end{eqnarray}
and then
the source and sink vector $\vec{S}$ is
\begin{eqnarray}
    \vec{S} = \left[0,R^\nu_{Y_e}, -\rho \frac{d\phi}{dr} + \frac{2P}{r}, -\rho v_1 \frac{d\phi}{dr}  \right].
\end{eqnarray}

\subsection{Neutrino transport}\label{subsec:neutrino_trans}
{\tt GR1D} calculates neutrino transport with the M1 scheme with multi-energy groups~\citep{2011PThPh.125.1255S, 2013PhRvD..87j3004C, 2015ApJS..219...24O}. The M1 scheme is the approximate method that solves the Boltzmann equation up to the first two moments and employs an analytic closure for closing equations.
Interactions between neutrino and matter are calculated in advance as an opacity table with {\tt Nulib}~\citep{nulib}.
Table~\ref{tab:nu_interations} summarizes interactions used in our simulation.
The energy groups are logarithmically divided into 18 energies. The center energy of the lowest energy group is 2.0 MeV, and that of the highest energy group is 150 MeV because the temperature in WDs is so low that neutrinos are hardly produced in higher temperatures. 
In our simulations, neutrino transport is calculated out to 600 km.

\begin{table*}
    \centering
    \caption{Summary of neutrino-matter interactions. Here, $n$ is a neutron, $p$ is a proton, $(A, Z)$ is a nucleus whose mass number is $A$ and atomic number is $Z$. The neutrino reaction represented by the symbol $\nu$ is flavor-insensitive, while the reaction represented by $\nu_i$ is flavor-sensitive.}
    \label{tab:nu_interations}
    \begin{tabular}{cc}
        \hline \hline
        Neutrino productions & References\\
          \hline
         $\nu_e + n \rightarrow p + e^- $ & \cite{Burrows_2006,Horowitz:2001xf}\\
         $\bar{\nu}_e + p \rightarrow n + e^+$ &  \cite{Burrows_2006}\\
         $\nu_e + (A,Z) \rightarrow (A, Z+1) + e^-$ & \cite{Burrows_2006, Bruenn:1985en}\\
         $e^- + e^+ \rightarrow \nu_x + \bar{\nu}_x$ & \cite{Burrows_2006, Bruenn:1985en} \\
         $N + N \rightarrow N + N +\nu_x + \bar{\nu}_x$ &  \cite{Burrows_2006, Bruenn:1985en} \\
         \hline
         Neutrino scattering \\
         \hline
         $\nu + \alpha \rightarrow \nu + \alpha$&\cite{Burrows_2006, Bruenn:1985en} \\
         $\nu_{ i} + p \rightarrow \nu_{ i} + p$&\cite{Burrows_2006, Bruenn:1985en,Horowitz:2001xf}  \\
         $\nu_{ i} + n \rightarrow \nu_{ i} + n$&\cite{Burrows_2006, Bruenn:1985en,Horowitz:2001xf}  \\
         $\nu_{ i} + (A,Z) \rightarrow \nu_{\rm i} + (A,Z)$&\cite{Burrows_2006, Bruenn:1985en, Horowitz_1997}  \\
         $\nu_{i} + e^- \rightarrow \nu^\prime_{i} + e^{-\prime}$&\cite{Bruenn:1985en,1994ApJ...433..250C}\\
         \hline \hline
         
    \end{tabular}
\end{table*}

\subsection{Equation of state}\label{sub:EoS}

We make a new Equation of State (EoS) table for {\tt GR1D}. We use EOSmaker\citep{2010CQGra..27k4103O, eos_maker}, which makes EoS tables with extrapolation, interpolation, and connection between different tables.
We adopt the H-Shen EoS \citep{Shen:2020sec} above density of $10^{10}{\,{\rm g\,cm^{-3}}}$.
The H-Shen EoS is the relativistic mean-field model, consistent with nuclear experiments and NS observations, and has a small symmetry energy slope $L=40$ MeV. We use the Timmes EoS \citep{Timmes_2000} as an EoS table in the whole region for electrons and photons.
To smoothly connect the H-SHen EoS and the Timmes EoS at the density of $10^{10}\,{\rm g\,cm^{-3}}$, we assume an ideal gas composed of electrons, photons, neutrons, protons alpha particles and heavy nuclei whose average $A$ and $Z$ are given by the H-Shen EoS.

During WD simulations, specific internal energy, density, and temperature go out of the EoS table, and simulation stops, especially in low-density regions. Thus, for specific internal energy and density, we impose lower limits on the values of the atmosphere. The atmosphere of specific internal energy is $-2.21\times 10^{18}\,{\rm erg\,g^{-1}}$ and that of density is 2.0 ${\rm g~cm^{-3}}$. The specific internal energy can be negative because we have to consider the binding energy of nuclei. For temperature, we do not directly change the temperature during simulations. Alternatively, when the simulation stops, we increase entropy up to $2~k_{B}/{\rm baryon}$ to increase temperature.
The reason why we change entropy is that it is easier than to change temperature directly.  The prescription is justified because the atmosphere density is low enough. We tried 1.5$k_{\rm B}$/baryon, 2$k_{\rm B}$/baryon and 2.5$k_{\rm B}$/baryon. The increase of 1.5$k_{\rm B}$/baryon did not prevent temperature from going down the lower limit. The increases of 2.0$k_{\rm B}$/baryon and 2.5$k_{\rm B}$/baryon prevented temperature from going out of the EoS table and we obtained the same fluid results. Thus, we adopted 2.0$k_{\rm B}$/baryon.
The prescription to entropy is needed for Model4.0. Temperatures of model1.0 and model2.0 did not go down below 0.05MeV.
WD profiles are prepared out to about 2,000 km, and our simulation region is 10,000 km. We calculate the mass of the atmosphere and get $2.0\times10^{-6}\,M_\odot$, which is much smaller than the ejecta mass.

\section{Results}\label{sec:results}
This section provides the results of our AIC simulation in order. A WD star is initially in hydrostatic equilibrium, and this is confirmed in Section~\ref{subsec:adiabatic}. 
Then, as electron capture occurs at the center, the central electron fraction decreases with time, removing degenerate electrons and reducing pressure. This induces a collapse toward the center of the star (collapse phase, see section~\ref{subsec:collapse}). 
Eventually, when the nuclear density is achieved, material bounces off the center, creating a shock wave, and neutrino heating from the center accelerates the shock wave and ejects the outermost layers (explosion phase, see section \ref{subsec:explosion}). 
The main goal of this paper is to clarify the physical conditions under which a WD undergoes collapsing and exploding via the electron capture reaction: 
In Sections~\ref{subsec:adiabatic} and \ref{subsec:collapse}, we discuss results for WDs with a total mass of $1.6M_\odot$ and different initial central densities $10^{9}-10^{10}$\,g\,cm$^{-3}$, and in Section~\ref{subsec:explosion}, we report on the explosion process of Model2.0 as a representative example.

\subsection{Simulations without weak interactions}\label{subsec:adiabatic}
\begin{figure}
    \centering
    \includegraphics[width=0.49\textwidth]{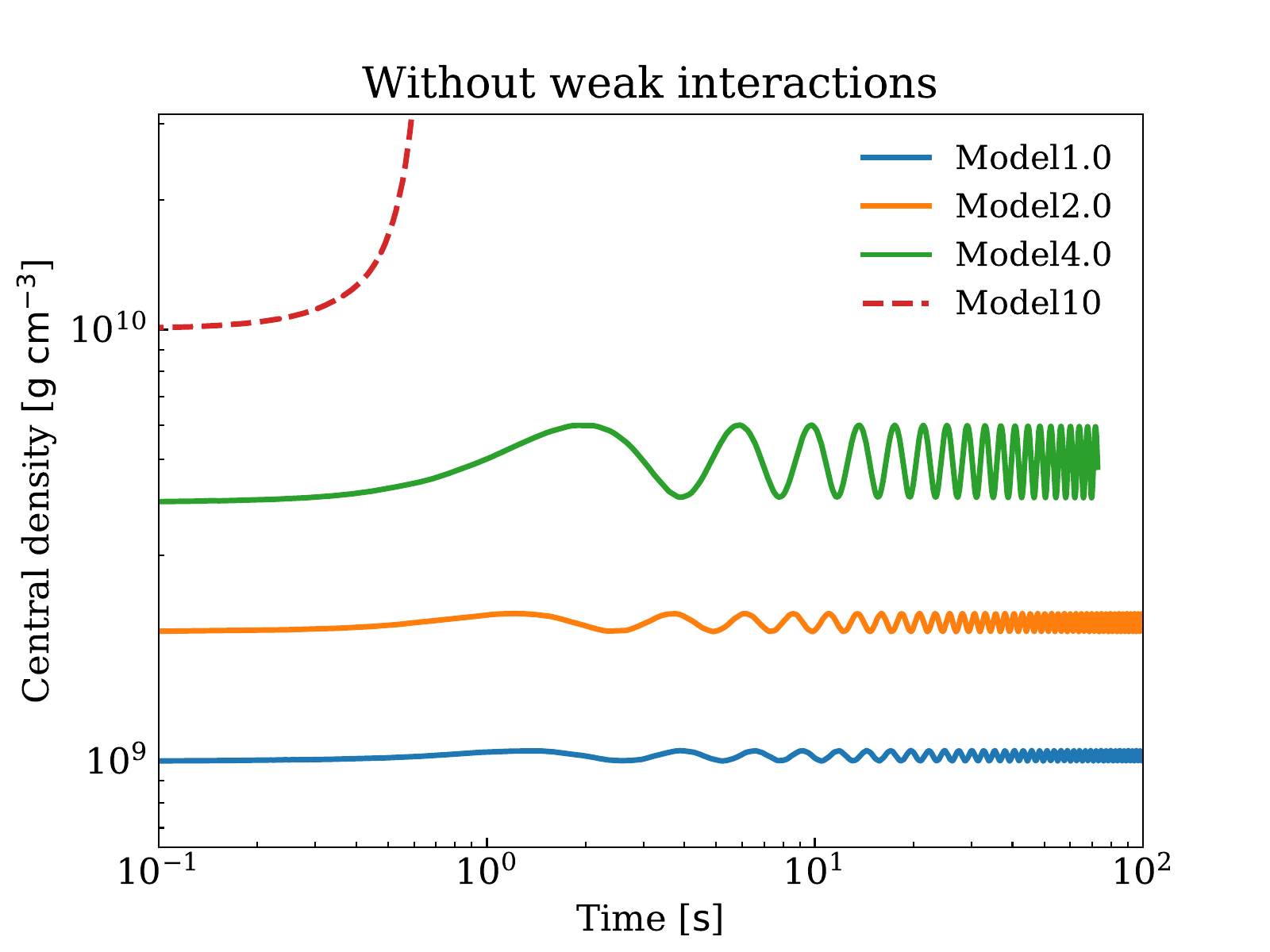}\caption{Time evolution of the central densities without weak interactions in the Newtonian gravity. The rapid increase shows the $1.0\times10^{10}{\rm\,g\,cm^{-3}}$ model (dashed line) collapses. There are stable numerical pulsations in the solid lines.}
\label{fig:rhoc_ad}
\end{figure}

Figure~\ref{fig:rhoc_ad} shows the time evolution of the central densities in an environment without weak interactions for the initial condition of WD.
We can first confirm that WDs with initial central densities $1.0\times10^{9}-2.0\times10^{9}{~{\rm\,g\,cm^{-3}}}$ are stable on a sufficient time scale and do not undergo any collapsing without the electron capture reaction. 
The pulsations in these WDs are caused by numerical effects: for example, a difference in addressing gravity and a mismatch of meshes. However, the amplitudes of the pulsations keep constant.
By contrast, Model10, whose initial density is $1.0\times10^{10}\,{\rm\,g\,cm^{-3}}$, is found to undergo the core-collapse in about 1 second due to photodisintegration reactions, which are taken into account in EoS.
Therefore, we use Model1.0, Model2.0 and Model4.0 (and Model2.0 as the standard model) for the collapse and explosion calculations that take electron capture reactions into account in the next section and hereafter.

\subsection{Collapse phase including weak interactions}\label{subsec:collapse}
\begin{figure}
\centering
  \includegraphics[width=0.49\textwidth]{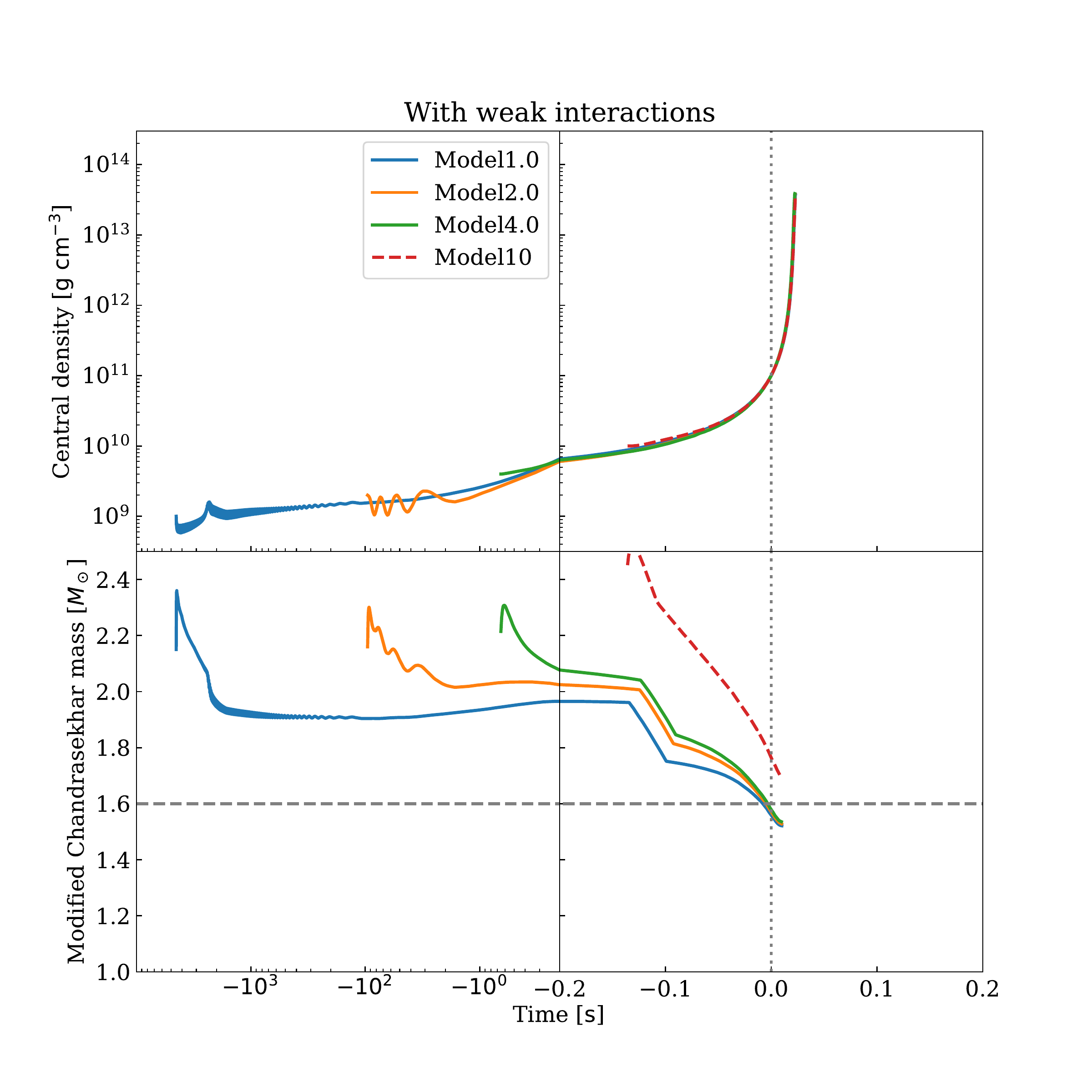}
    \caption{(top) Time evolution of the central densities of the WD when the weak interactions are taken into account, and (bottom) comparison of the WD mass with the modified Chandrasekhar masses estimated from the central physical quantity $(Y_{e,c}~\mathrm{and}~ s_{e,c})$ by Eq.~\ref{eq:Mch}. The origin of time is defined as the moment central densities reach $1.0\times10^{11}{\rm\,g\,cm^{-3}}$.
    In the two panels, the colors correspond to the initial central density models and the horizontal gray dashed line of the bottom panel corresponds to the total mass of the WD of $1.6 M_\odot$.}
\label{fig:rhoc_cc}
\end{figure}

To discuss the conditions for core collapse and the effect of the electron capture reaction, we show the time evolution of the WD mass and the critical mass above which the core becomes unstable in Figure~\ref{fig:rhoc_cc} (bottom).
As the critical mass, we introduce the equilibrium polytropic sphere plus the electron degenerate pressure and the finite temperature corrections (hereafter referred to as the modified Chandrasekhar mass; e.g., \cite{1990ApJ...353..597B}), following the modeling of \citet{2018MNRAS.481.3305S}.
We adopt the modified Chandrasekhar mass as
\begin{eqnarray}
    M_{\rm ch} &= 1.09M_\odot \left(\frac{Y_{e,c}}{0.42}\right)^{2} \times\label{eq:Mch} \\ \nonumber
    & \left[1+\frac{2}{3}\left(\frac{s_{e,c}}{\pi Y_e,c}\right)^2- 0.057 + \frac{s_{e,c}}{A}\right]^{3/2}
\end{eqnarray}
where $Y_{e,c}$ and $s_{e,c}$ are the central electron fraction and the central electronic entropy, respectively and $A$ is equal to 56 for our models.
average of $x$. We solved the Lane-emden equation of $n=3.3$ to derive the modified Chandrasekhar mass. If we use $n=3$ to solve the Lane-emden equation and set $s_{e,c}=0$ and $Y_{e,c}=0.5$, we get the usual Chandrasekhar mass $M_{\rm ch}=1.44M_\odot$.
Appendix~\ref{appendix} describes the derivation and see the appendix for the details.

Figure~\ref{fig:rhoc_ad} shows time evolution of the central densities without weak interaction. Model1.0, Model2.0 and Model4.0 keep stable for 1 minute. On the other hand, Model10 collapses in 1 second. The result implies another mechanism (e.g. photodissociation) makes WDs collapse. There are oscillations in  Model1.0, Model2.0 and Model4.0. These oscillations are eigenoscillations of models and triggered by mismatch of grids between initial models and the core-collapse simulations. About Model10, the time of the collapse is shorter than the period of the oscillation.

Figure~\ref{fig:rhoc_cc} (top) shows the time evolution of the central densities induced by the electron capture reaction of the WD star and up to the bounce. The origin of time is the moment when central densities reach $1.0\times10^{11}{\rm\,g\,cm^{-3}}$
We find that the standard model of Model2.0 undergoes core-collapse on a time scale of about $10$\,s, which is shorter than the time scale in Section~\ref{subsec:adiabatic}.
The inflection points for model 1.0, 2.0 and 4.0 around -0.12\,s due to the interpolation of the Timmes EOS and the HShen EOS. We find the same oscillations as Figure~\ref{fig:rhoc_ad}. The reason there is no oscillation in Model4.0 (green) is that the time scale of collapse of the green is faster that the periode of the oscillation.
The bump coincides with temporal decrease of the modified Chandrasekhar mass around 200 seconds. At that time, the electron fraction rapidly decreases. Numerical errors are likely to contribute to the bump, as the calculation runs for a few billion steps until the occurrence of the bump. Errors in the electron fraction propagate to other values, such as entropy, leading to a temporary change in the modified Chandrasekhar mass, ultimately resulting in the observed bump. However, this behavior does not impact the overall calculation, as the core collapse proceeds freely once a WD loses stability.
We can also confirm that the core collapse occurs on a shorter timescale for the higher initial central density model of Model4.0 and on a longer timescale for the lower initial central density model Model1.0.
Here we discuss two points: (1) this core collapse is actually due to electron capture reactions, and (2) the effect of the initial central density on the density structure evolution other than the core-collapse timescale.

Equation \ref{eq:Mch} gives the critical mass, above which the central density/pressure cannot support self-gravity anymore, and we can see that this mass falls as the central electron fraction decreases. The central electron entropy increases due to the heat the electron capture reaction produces.
We can see that as soon as the estimated modified Chandrasekhar mass of the blue/orange/green line reaches the WD mass in Figure~\ref{fig:rhoc_cc} (bottom), its central density increases sharply in Figure~\ref{fig:rhoc_cc} (top), i.e., the core-collapses.
It is well understood that this is not another instability but a physical collapse due to electron capture reactions.
On the other hand, Model10, indicated by the red line, shows a sudden increase in the central density at a stage where the modified Chandrasekhar mass that can be supported is sufficiently larger than the WD mass. This is due to instability in the initial conditions themselves, not due to electron capture. From the above, in this paper, we adopt Model1.0, Model2.0 and Model4.0 as the super Chandrasekhar mass WD models that have undergone appropriate gravitational collapse due to electron capture.

\begin{figure}
    \centering
    \includegraphics[width=0.49\textwidth]{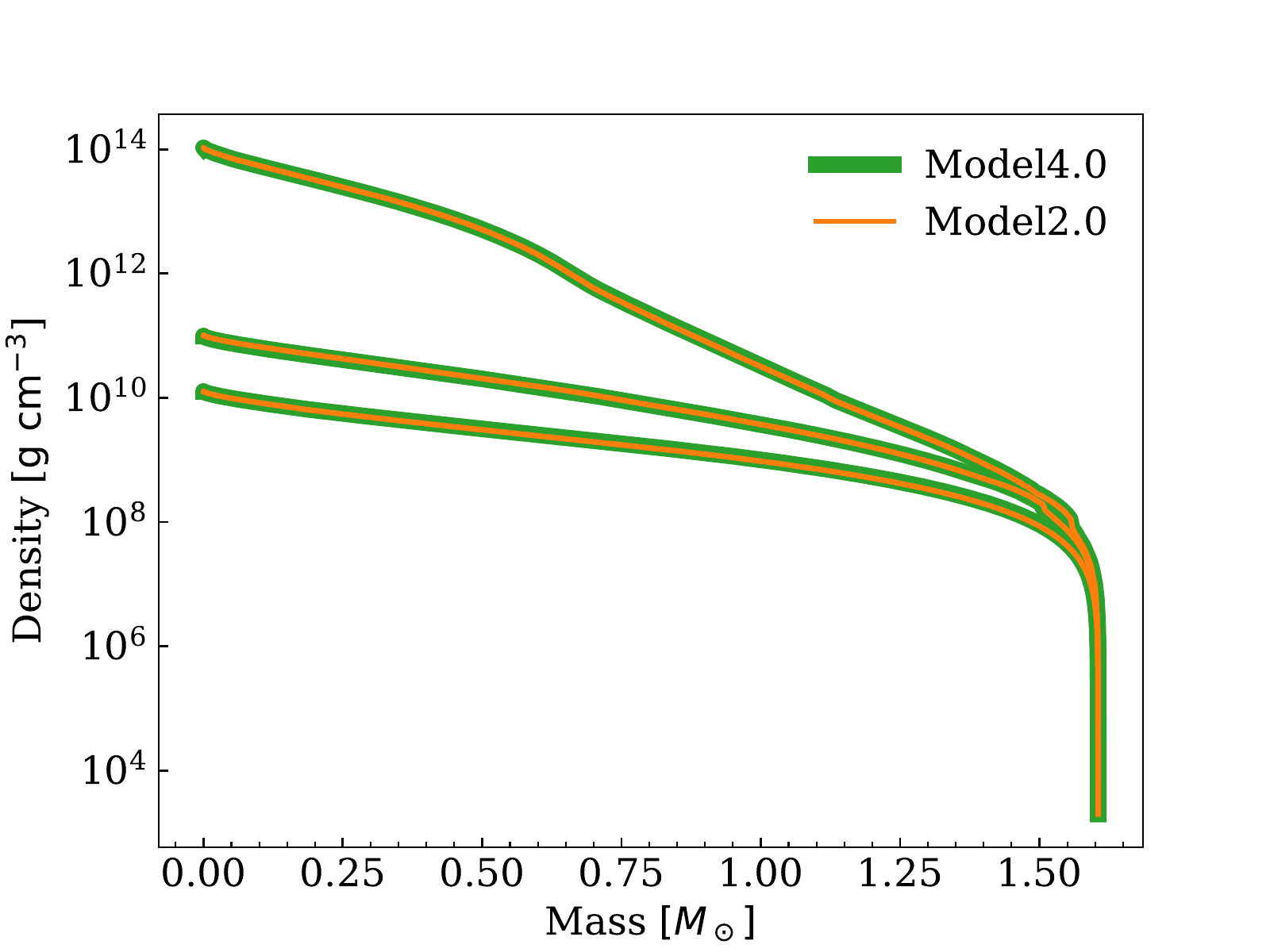}
    \caption{Snapshots of density structure evolution for Model4.0 (green) with different initial central densities indicated by different colors, when the central density reaches $10^{10}$, $10^{11}$, and $10^{14}$\,g\,cm$^{-3}$.}
    \label{fig:density_profile}
\end{figure}

Figure~\ref{fig:density_profile} illustrates snapshots of the density structure evolution for Model2.0 (orange) and Model4.0 (green) with different initial central densities when the central density reaches $1.0\times10^{10}$, $1.0\times10^{11}$, and $1.0\times10^{14}$\,g\,cm$^{-3}$.
This figure clearly reveals that even from different initial central densities, a WD follows approximately the same density-structure evolution after the core collapse due to electron capture reactions.
This result suggests that gravitational collapse proceeds along the same path regardless of the central densities.

\subsection{Explosion phase}\label{subsec:explosion}
\begin{figure}
    \centering
    \includegraphics[width=0.49\textwidth]{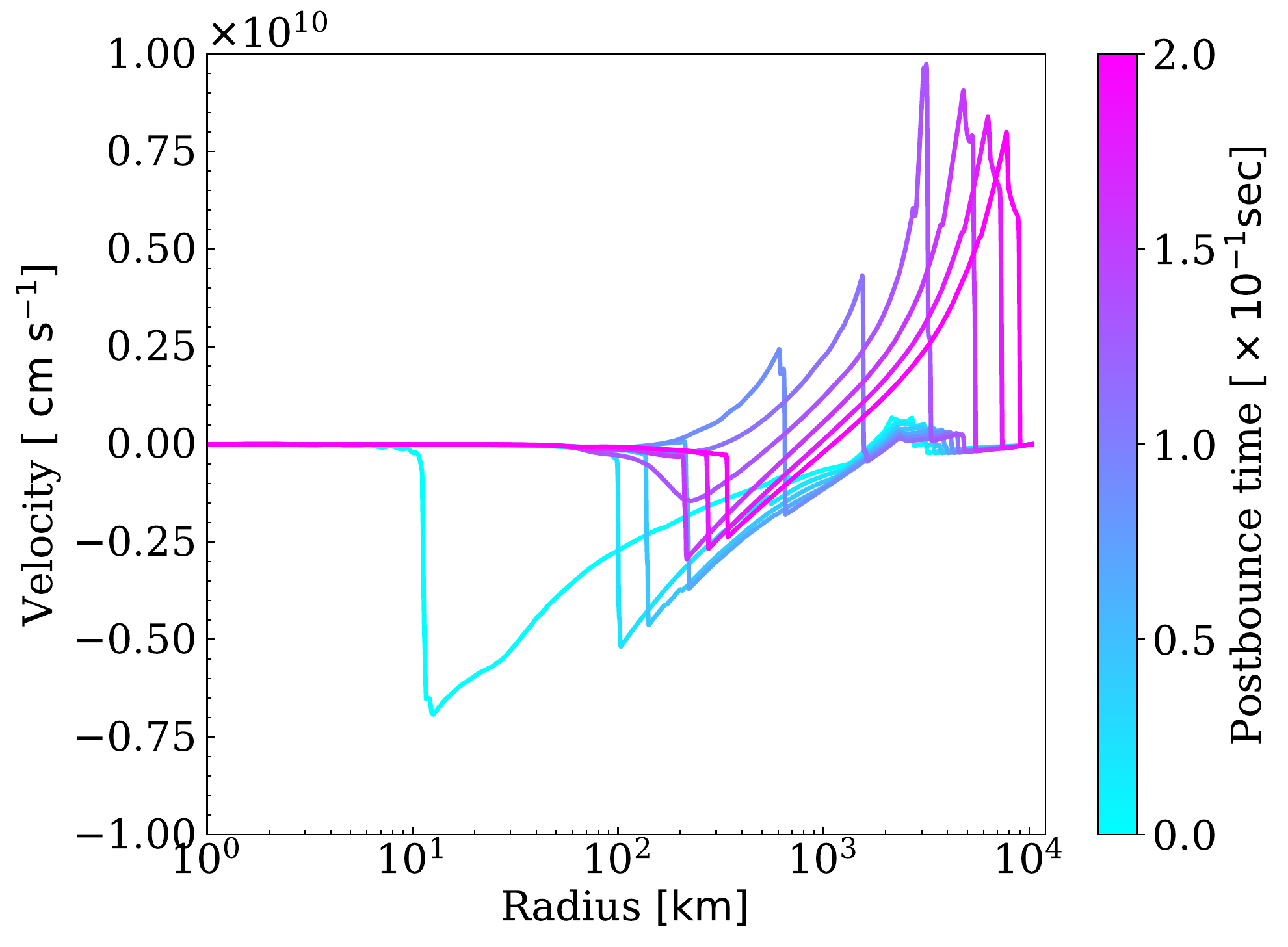}
    \caption{Velocity profiles after the bounce, each snapshot time corresponds to approximately every $20.0$ ms from the bounce times to after $0.2$ seconds. This illustrates Model2.0.}
    \label{fig:velocity_profile}
\end{figure}

Figure~\ref{fig:velocity_profile} shows the radial velocity profiles after the core bounce.
We treat an example of Model2.0 throughout this section.  We change to the general relativity when central densities reach $1.0\times10^{10}{\rm\, g\,cm^{-3}}$.
As shown in the figure, just after the bounce, a shock wave is formed around 10 km, with material accreting outside of the shock wave.
This shock radius gradually moves outward with time, and the velocity at the edge of the shock wave turns positive when it exceeds 300 km. The shock wave then propagates outward, confirming a successful explosion.

\begin{figure}
    \centering
    \includegraphics[width=0.49\textwidth]{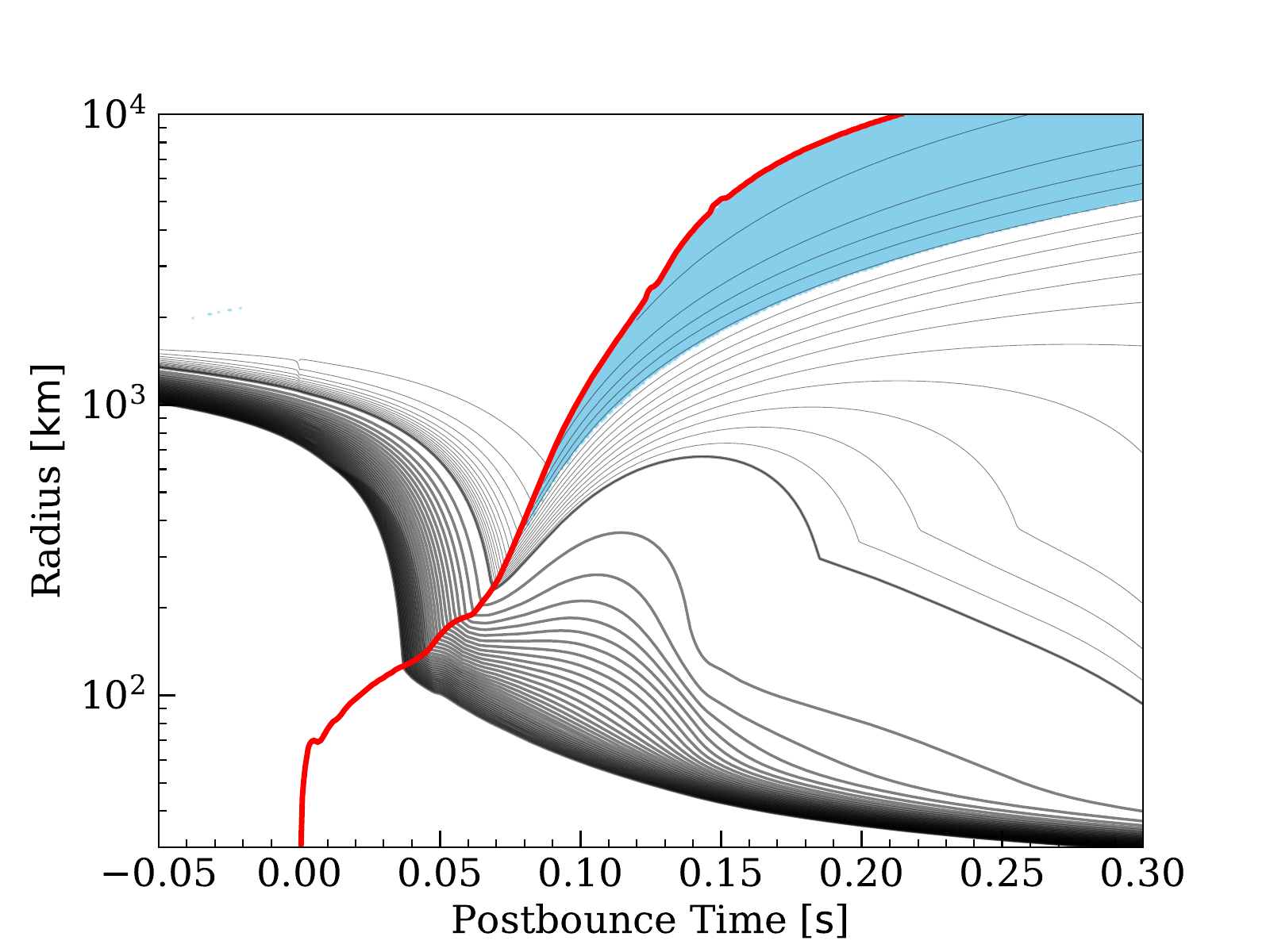}
    \caption{Radius evolution of Lagrangian mass coordinates with time, and the time evolution of the shock radius.
    This illustrates Model2.0. 
    The thick black solid lines are the mass shells, spaced in steps of $10^{-3}\,M_\odot$, and the thin black lines are spaced $10^{-4}\,M_\odot$. The minimum mass shell surrounds $1.57M_\odot$ and the maximum mass shell surrounds $1.605M_\odot$. The red line marks the shock radius of the explosion model. The blue indicates the region which meets Eq.~\ref{eq:escape-condition}.}
    \label{fig:ejecta_mass}
\end{figure}

Figure~\ref{fig:ejecta_mass} depicts mass shell trajectories as a function of time of the $1.0\times10^{9}$\,g\,cm$^{-3}$ explosion model. 
After the bounce, the formed shock wave is stalled around $\sim200$\,km until 0.07 seconds, and then, the shock wave penetrates the outermost layers.
The blue-filled region shows the ejecta component that satisfies the following escape conditions:
\begin{eqnarray}
    -(1-\alpha) + \frac{1}{2}v^2 + \epsilon > 0 \label{eq:escape-condition}~,
\end{eqnarray}
and these mass is estimated to be $M_\mathrm{ej}\sim 5\times 10^{-4}\,M_\odot$. The criterion is similar to the Bernoulli criterion. However, our criterion is the Newtonian formula and more strict than the Bernoulli criterion \citep{2023CQGra..40h5008H}. We should consider the ejecta mass as the lower limit.
Since the shock wave takes a maximum velocity of about $\sim10^{10}$ cm s$^{-1}$ (see Figure \ref{fig:velocity_profile}) immediately after penetrating the outermost layer of the WD, the ejecta of this explosion is expected to have a very small mass moving at a much larger velocity of nearly $\sim0.1c$.

It should be noted that in the figure, there is a bending behavior in the trajectory of the material that failed to escape at $\sim300$ km after $0.15$ seconds. This is not due to a numerical error but to secondary shock waves generated by the fallback accretion of material on the central PNSs.
The bending itself is due to the nature of the WD because the atmosphere is thin enough. 
However, the behavior of this fallback material and the time evolution of the shock wave velocity of the ejected material depend on the treatment of the atmosphere density in this calculation, which will be discussed in Section \ref{sec:discus}.

\begin{figure}
    \centering
    \includegraphics[width=0.49\textwidth]{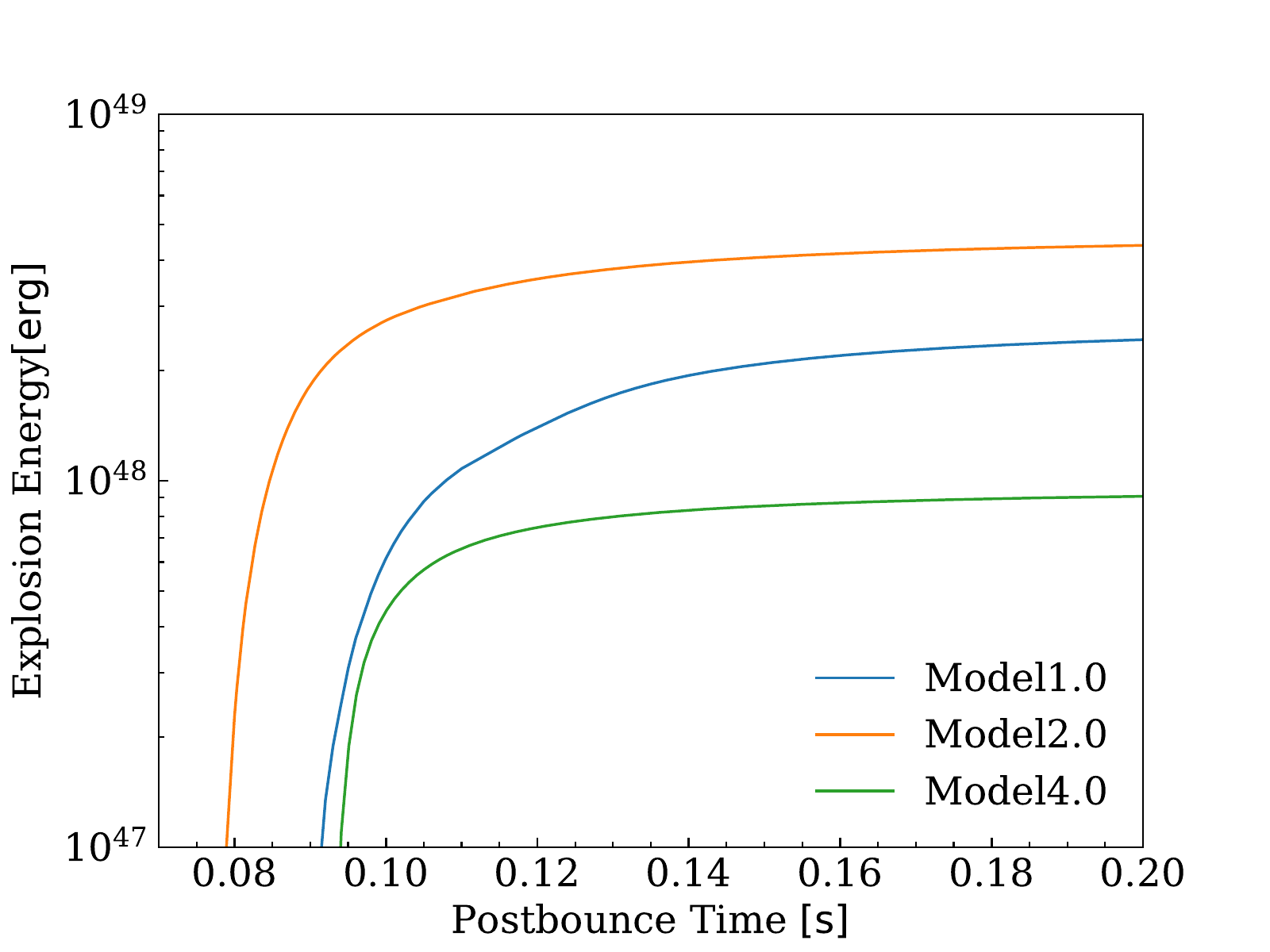}
    \caption{The time evolution of the explosion energy}
    \label{fig:explosion_energy}
\end{figure}

Figure~\ref{fig:explosion_energy} also shows the explosion energy as a function of time. The explosion energy is estimated by integrating the total energy of elements of material that satisfy the escape conditions of Eq.\,\ref{eq:escape-condition} and writes 
\begin{eqnarray}
    E_{\rm expl} &= \int_{e_{\rm bind} > 0}e_{\rm bind}d\tilde{V}, \label{eq:energy-definition}\\
    e_{\rm bind} &= \alpha(\rho (1 + \epsilon + P/\rho)W^2 - P) - \rho W,
\end{eqnarray}
where $d\tilde{V}$ is the three-volume element for the curved space-time metric~\citep{2012ApJ...756...84M}.
The explosion energy reaches about $\sim$ a few$\times10^{48}$ ergs and converges almost immediately after the bounce.
The results, including Model1.0 and Model4.0 are summarized in Table~\ref{tab:summary-result}. 
While, as noted in Figure \ref{fig:density_profile}, it is suggested that all models of Model1.0, Model2.0 and Model4.0 follow approximately the same gravitational collapse process and density structure evolution, some variation in the resulting explosion energies and ejecta masses is confirmed.
However, the binding energy of the hot outer layer that remains around the WD as a result of the dynamical merger of the two WDs is $E_\mathrm{env,bind} \sim GM_\mathrm{env}M_\mathrm{WD}/R_\mathrm{env}\sim10^{48}\,\mathrm{erg}$, where we assumed that the cooling timescale of the outer layer is sufficiently long for the collapse timescale and $M_\mathrm{env}\sim0.1M_\odot$ \citep{2016MNRAS.463.3461S}. 
This implies the explosion's outcomes are strongly dependent on the outer layer structure.
For comparison, we also show the result of the 1.5$M_\odot$ WD, whose central density is $2.0\times10^{9}\,{\rm g\,cm^{-3}}$. The ejecta mass and the explosion energy are two orders of magnitude smaller than those of 1.6$M_\odot$ WDs.
Furthermore, our core-collapse simulation is the 1D spherical symmetry. Please note that it is suggested that the explosion energy may increase due to multidimensional effects (e.g., \cite{2015ApJ...801L..24M}).

\section{Discussion}\label{sec:discus}

\begin{figure}
    \centering
    \includegraphics[width=0.49\textwidth]{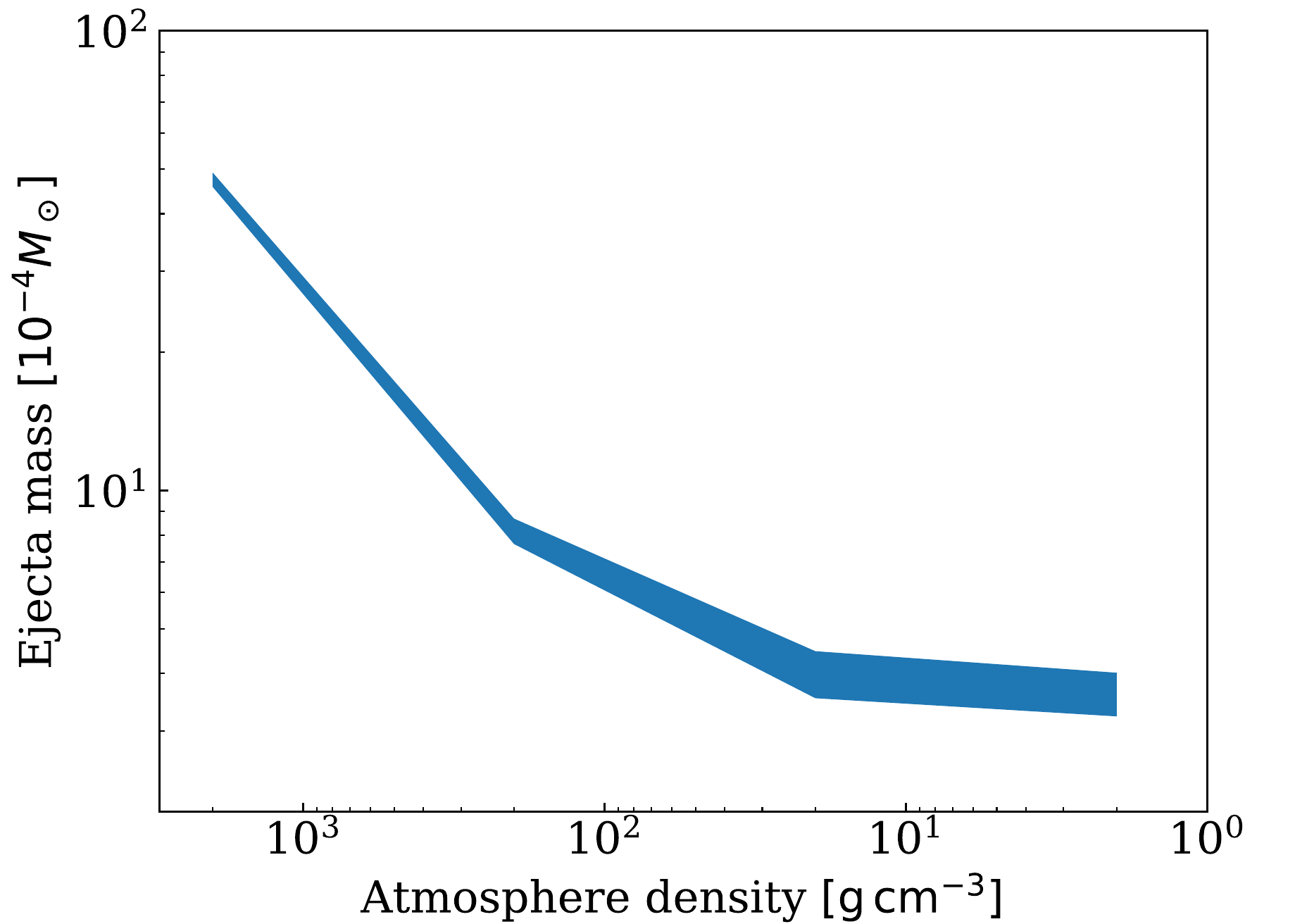}
    \caption{Ejecta mass distribution that is measured between 0.1\,s and 0.3\,s regarding different atmosphere densities for Model2.0. The horizontal axis is atmosphere density, and the right side is smaller. The vertical axis is ejecta mass in $10^{-4}M_\odot$.} 
    \label{fig:ejecta_mass_93}
\end{figure}

 We discuss a few caveats in the following.
In GR simulations, we have used an atmosphere outside WDs to avoid a low-density environment outside the EoS table as Section~\ref{sub:EoS}. We run additional simulations, in which the density of the atmospheres are changed for Model2.0: 20\,$\rm g\,cm^{-3}$, 200\,$\rm g\,cm^{-3}$ and 2,000\,$\rm g\,cm^{-3}$.
Figure~\ref{fig:ejecta_mass_93} shows ejecta masses from 0.1\,s to 0.3\,s with respect to atmosphere densities. Ejecta masses increase as atmosphere densities increase because the shock wave must sweep out heavier masses, and ejecta masses converge below 20\,$\rm g\,cm^{-3}$. Hence our density atmosphere of 2.0\,$\rm g\,cm^{-3}$ dose not affect explosion. 

\begin{table}[]
    \centering
    \caption{Summary of simulation results. Ejecta masses and explosion energies are both measured at 0.2\,s in postbounce time. Atmosphere densities are set $2.0\,{\rm g\,cm^{-3}}$} 
    \begin{tabular}{ccc|ccc}
    \hline\hline
    Mass & Model &initial $\rho_c$ & $M_\mathrm{ej}$ & $E_\mathrm{expl}$ \\
   \lbrack $M_\odot$\rbrack & & \lbrack cm$^{-3}$\rbrack  & \lbrack $10^{-4}M_\odot$\rbrack & \lbrack $\rm 10^{48}erg$\rbrack \\ \hline
    \multirow{3}{*}{1.6} & Model4.0 &$4.0\times10^{9}$ & $1.1$ & $0.9$\\
     & Model2.0& $2.0\times10^{9}$ & $4.4$ & $4.3$\\
     &Model1.0& $1.0\times10^{9}$ & $2.7$ & $2.4$ \\
     \hline
     1.5 & &$2.0\times10^{9}$ & $0.45$& 0.046\\
    \hline\hline
    \end{tabular}
    \label{tab:summary-result}
\end{table}

Table~\ref{tab:summary-result} summarizes ejecta masses and explosion energies of our calculations of three $1.6 M_\odot$ WDs and one $1.5M_\odot$ WD. 
Table~\ref{tab:summary-result} indicates that the explosions are weak with ejecta masses of $10^{-5}-10^{-4}M_\odot$ and explosion energies being $10^{46} - 10^{48}\,{\rm erg}$.
The atmosphere densities are set $2.0{\rm \,g\,cm^{-3}}$ for all the models in Table~\ref{tab:summary-result}. Hence, the ejecta masses are regarded as the minimum ejecta masses.

 Ejecta masses reported in previous studies are $10^{-3}M_\odot$~\citep{2006ApJ...644.1063D}, $10^{-2}-10^{-1} M_\odot$~\citep{2020ApJ...894..146S} and $10^{-3}M_\odot$~\citep{2023MNRAS.518..623M}. Compared to previous studies, our minimum values are one to two orders of magnitude smaller than ejecta masses of previous studies. The simulation by ~\cite{2006ApJ...644.1063D} is two-dimensional simulation but two latter simulations~\citep{2020ApJ...894..146S, 2023MNRAS.518..623M} are one-dimensional simulation same as ours. According to \cite{2023MNRAS.525.6359L}, they carried out three-dimensional simulations with the M1 scheme and general relativity and they reported there was no ejection for their non-rotating model. These previous studies imply that the treatment of neutrino and gravity may have a larger influence than dimension. Improvement in the gravity or (and) the neutrino transport is likely to explain the difference.

A recent theoretical model by \cite{2017ApJ...842...34W} suggested that the ejecta mass from FRB~121102~\citep{2014ApJ...790..101S,2016ApJ...833..177S} is $\sim10^{-5}M_\odot$ (see also \cite{2017ApJ...839L...3K}). Our ejecta mass is closer to the value, which may be realized by our simulations' improved treatment of gravity and neutrino. Note that the mass ejection is likely to be affected by structures of outer atmospheres in Figure~\ref{fig:ejecta_mass_93}.

In addition, for very small orders of ejecta mass, as in this case, the contribution of the neutrino-driven wind, which is the mass ejection from the PNS surface by neutrinos, should be discussed. 
Our calculations at the time of Figure~\ref{fig:ejecta_mass_93} indicate that the mass flux from the PNS surface is $\sim10^{-4}M_\odot$ s$^{-1}$.
Our calculations also suggest that the luminosity of neutrinos in the late phase decays from the order of $L_\nu\sim10^{51}$\,erg\,s$^{-1}$ for all flavors. The average energy of neutrinos is $\sim$10\,MeV, which agrees with another imporved neutrino calculation and is the same as the typical value of normal core-collapse supernovae since the structure of the PNSs is similar.

\begin{figure}
    \centering
    \includegraphics[width=0.49\textwidth]{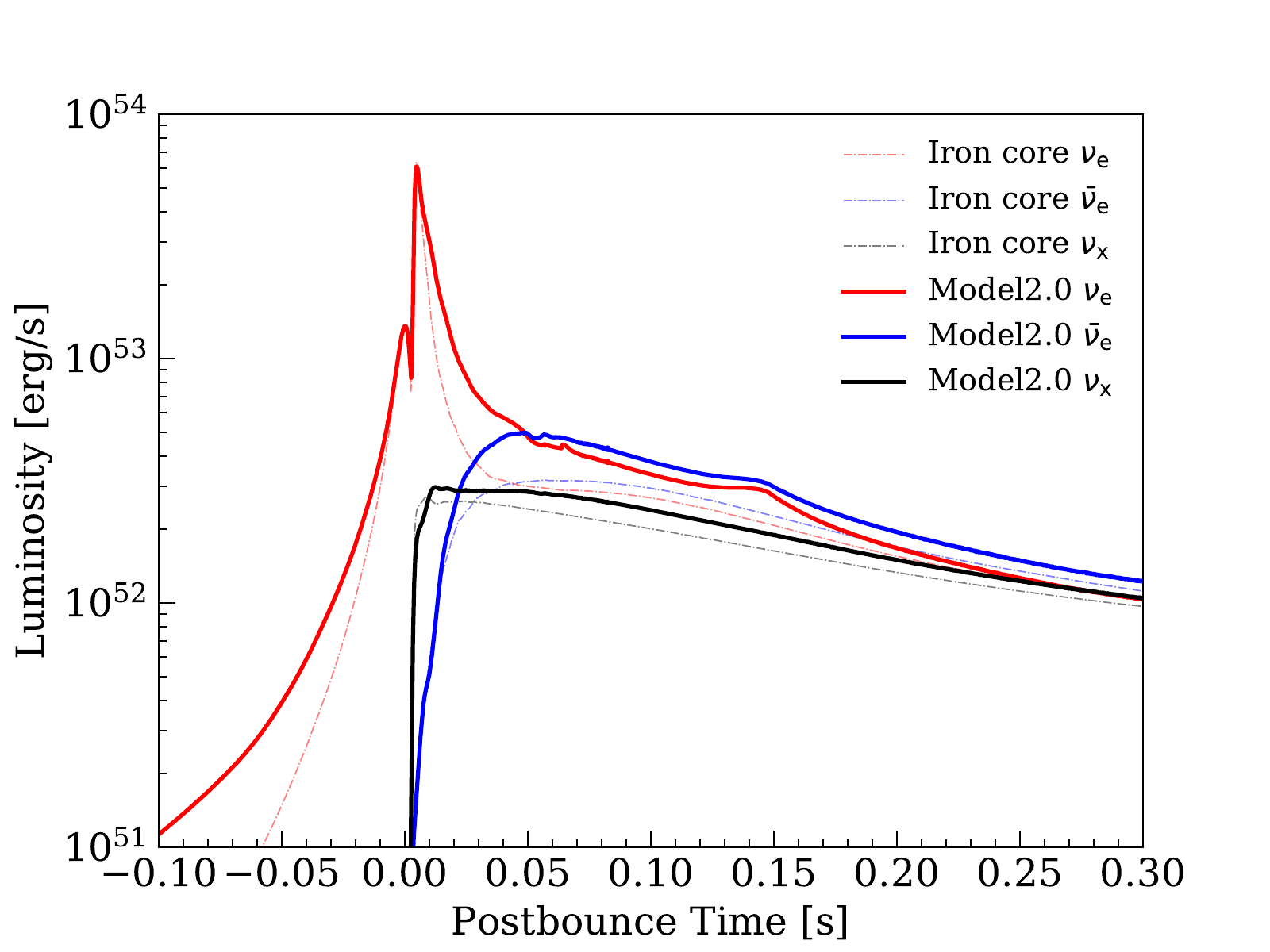}
    \includegraphics[width=0.49\textwidth]{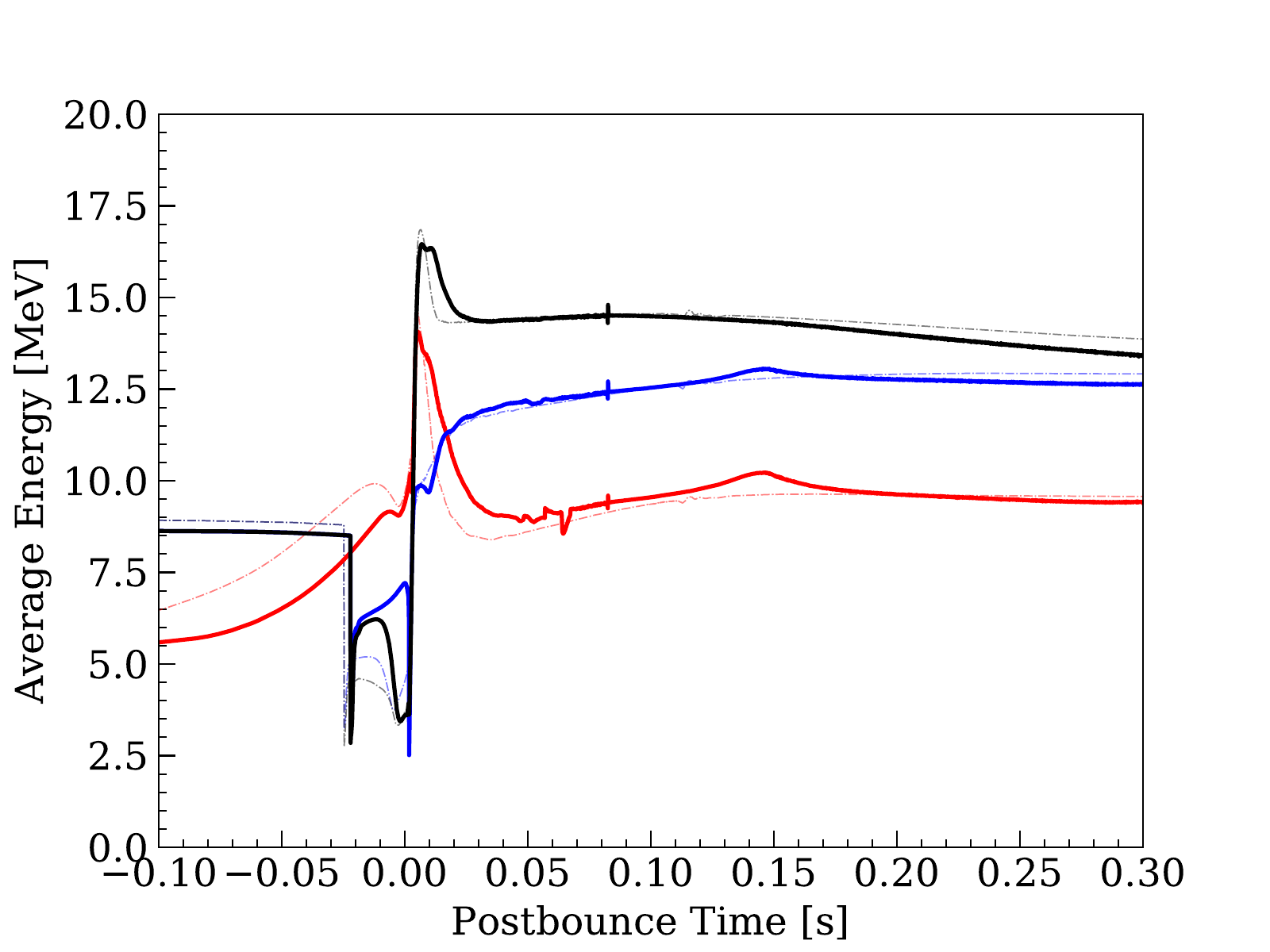}
    \caption{The time evolution of neutrino luminosity (top) and average energy (bottom) for Model2.0. The red is eletron neutrino, the blue is anti-electron neutrino and the black is heavy-lepton neutrino. For comparison, the result of the core collapse of the iron core~\citep{2021PTEP.2021b3E01M,2023PhRvD.107h3015M} is also shown in the dashed lines.}
    \label{fig:luminosity_93}
\end{figure}

Figure~\ref{fig:luminosity_93} shows time evolution of neutrinos for Model2.0. Luminosity of electron neutrino is $10^{51}{\rm\,erg\,s^{-1}}$ at -0.1\,s and increases. There is a dip in the electron neutrino luminosity at the bounce. This is due to so-called neutrino trapping. In the aftermath of the dip, the electron neutrino luminosity takes a peak of $6\times 10^{53}{\,\rm erg\,s^{-1}}$. The luminosities of anti-electron neutrino and heavy-electron neutrino rapidly increases after the bounce. The highest values are $5\times10^{52}{\rm\,erg\,s^{-1}}$ for anti-electron neutrino and $2\times10^{52}{\rm\,erg\,s^{-1}}$ for heavy-lepton neutrino. Then, they progressively decrease to $1\times10^{52}{\rm\,erg\,s^{-1}}$ for electron and heavy-lepton neutrinos $1\times10^{52}{\rm\,erg\,s^{-1}}$ and $1.5\times10^{52}{\rm\,erg\,s^{-1}}$ for anti-electron neutrino at 0.3\,s. There are bulges in electron and anit-electron neutrinos at 0.15\,s. This is due to the temporary failure of explosion from 0.15\,s in Figure~\ref{fig:ejecta_mass}.

About average energies, the average energy of electron neutrino has the peak of 15\,MeV at the bounce, which decays for 0.03\,s to 10\,MeV, and keeps 10\,MeV until 0.3\,s. The average energy of heavy-lepton neutrino has the peak of 17.4\,MeV, which decays for 0.02\,s to 15\,MeV, and keeps 15\,MeV until 0.3\,s. The average energy of anti-electron neutrino is 10\,Mev and increase to 12.5\,MeV.

Comparing to the explosive 1D core-collapse of the iron core \citep{2021PTEP.2021b3E01M,2023PhRvD.107h3015M}, the neutrino luminosity of Model2.0 is from 1.5 to 2.0 higher than that of the core collapse of the iron core. The difference is due to the neutron star masses after explosions of the AIC ($1.6M_{\odot}$) and the iron core ($1.36M_\odot$). the average energies have the same shapes as those of the core collapse of the iron core. However, there are bulges in the luminosity and the average energy of electron and anti-electron neutrinos. This is due to the temporary shrink of the shockwave in Figure~\ref{fig:ejecta_mass_93}. Moreover, we compare to the previous AIC studies~\citep{2006ApJ...644.1063D,2023MNRAS.525.6359L} and find out that the luminosities of our model have the same orders as those of both previous studies, which are from $10^{52}{\rm\,erg\,s^{-1}}$ to $10^{53}{\rm\,erg\,s^{-1}}$.

From the neutrino luminosity and PNS radius in this simulation using the analytical neutrino-driven wind model \citep{1996ApJ...471..331Q,2021ApJ...908....6S}, the mass ejection rate is calculated to be $\sim10^{-4}M_\odot$ s$^{-1}$, which is roughly consistent with the mass flux at the surface. 
From this, even assuming a neutrino-driven wind of more than about 10 seconds, we can only expect an increase in the ejecta mass of the order of $\sim10^{-3}M_\odot$ at the highest.

We did not consider rotation and magnetic field. They have important roles in NSs. Both effects can help explosion but decrease neutrino luminosity and energy \cite{2005ApJ...620..861T}. They have negative and positive effects on ejecta masses.
Rotation seems to increase ejecta mass~\citep{2006ApJ...644.1063D}. They calculated slowly and rapidly rotating AICs, whose ratios of rotational energy to gravitational energy $T/|W|$ are 0.059 and 0.263, where $W$ is gravitational energy and $T$ is rotational energy. They showed that both ejecta masses are $\sim 10^{-3}M_\odot$. This value is covered with the distribution in Figure~\ref{fig:ejecta_mass_93}, which implies that the influence of rotation can increase ejecta mass but the uncertainty of rotation is comparative to that of surrounding environments.
In the next, we will consider both effects. GR1D can address rotation. About magnetic fields, we have to phenomenologically incorporate effects of magnetic field.

\section{Summary}\label{sec:summary}
In this paper, we have reported results of self-consistent simulations for WDs from their hydrostatic initial conditions, through collapse by electron capture reactions, to explosions and the formation of PNSs. It is important that we have checked the stability of initial models and used elaborate methods for the general relativistic neutrino-radiation hydrosimulation. We proved that the core-collapse of three models of $1.6M_\odot$ WDs whose central densities are $1.0\times10^{9}$, $2.0\times10^{9}$ and $4.0\times10^{9}{\rm\,g\,cm^{-3}}$ and a model with $2.0\times10^{9}{\rm\,g\,cm^{-3}}$ of $1.5M_\odot$ are indeed triggered by the electron capture and they cause weak explosions. In conclusion, the minimum ejecta masses are explosion energies are $\sim 1.0^{-4}M_\odot$ and $\sim10^{48}$\,erg for the $1.6M_\odot$ WDs and $10^{-5}M_\odot$ and $10^{46}$\,erg for the $1.5M_\odot$ WD. The ejecta masses increase with atmosphere densities. The improved treatment of neutrinos and gravity lead to the values. We determined the minimum ejecta masses from AIC of $1.6M_\odot$ WDs. Note that the results in this paper are for non-rotating case. In \cite{2023MNRAS.525.6359L}, they describe rotating models lead to more ejecta than the non-rotating model.

This paper has provided results of core-collapse and the explosion of 1.6$M_\odot$ WDs.
Although the mass of 1.6$M_\odot$ is heavy for WDs, this result is very important for discussing the gravitational collapse timescale of WD due to electron capture.
It takes a longer time to calculate the core-collapse of lighter WDs. For example, it takes one month and a half to calculate the core-collapse of a WD of $1.5M_\odot$ with 8 threads of Intel(R) Core(TM) i7-7820X CPU @ 3.60GHz. This length is a few times longer than that of $1.6M_\odot$.
When we assume the same rate to be kept, the calculation time of $1.4M_\odot$ should be half of a year and the calculation is unrealistic.
In the simulation, the speed of single-thread is more important than multi-thread performance. Even if we use more threads, it does not accelerate our simulations.

Figure~\ref{fig:density_profile} implies structures during core-collapse do not depend on initial models. The difference in ejecta masses is due to atmospheres. If we can prepare profiles during core-collapse for arbitrary masses, it saves us to calculate heavy long-term neutrino-radiation hydrodynamics. 
In future work, we plan to calculate $1.4M_\odot$ WDs for this method.


\section*{Acknowledgments}
M.M. and Y.S. thank K. Sumiyoshi for providing the nuclear EOS table.
This work has been supported by Japan Society for the
Promotion of Science (JSPS) KAKENHI grants 
(18H05437, 19K03907, 20H00174, 20H01901, 20H01904, 20H05852, 21J00825, 21K13964, 22H04571, 22KJ0528).
and by the NSF Grant No.~AST-1908689, No.~AST-2108466 and No.~AST-2108467.

\appendix
\section{The modified Chandrasekhar mass}\label{appendix}
Here, we discuss the conditions for core-collapse by comparing the WD mass to a critical mass above which the core becomes self-gravity unstable.
As the critical mass, we introduce the equilibrium polytropic sphere plus the electron degenerate pressure and the finite temperature corrections following the modeling of \citet{2018MNRAS.481.3305S}.
First, we approximate the WD core as a polytropic sphere in equilibrium. 
While the WD core is not strictly isoentropic, this approximation provides a good prediction of the profile of the WD core. 
The equilibrium configurations are expressed by the {\it Lane-Emden equation}
\begin{eqnarray}
    &\frac{1}{\xi^2}\frac{d}{d\xi}\left(\xi^2\frac{d\theta}{d\xi}\right)=-\theta^n\label{eq:LE1}~,
\end{eqnarray}
where we follow the standard notation of \citet{1967aits.book.....C} and $n$ is the polytropic index, $\theta$ is the dimensionless density and $\xi$ is the dimensionless radius.
Using the central density and pressure, the mass can be obtained as follows and $\theta$ and $\xi$ are dimensionless radius and density, respectively.
\begin{eqnarray}
    &M_{\rm ch} = \left(\frac{1}{4\pi G^3}\frac{P_c^3}{\rho_c^4}\right)^{1/2}\varphi_n \label{eq:LE2}~,\\
    &\varphi_n=(n+1)^{3/2}\left.\left(\xi^2\frac{d\theta}{d\xi}\right)\right|_{\theta(\xi_n)=0}
    \label{eq:LE3},
\end{eqnarray}
where $\varphi_n$ depends on $n$. For example, $n = 3.3$ gives $\varphi_n\approx17.27$. 
The physical meanings of these values are that $M$ is the dimensionless mass of a star and $\phi_n$ is the dimensionless maximum radius of a star.
Also, $\rho_c$ and $P_c$ are the central density and pressure and are related to each other as $P_c = K\rho_c^{1+1/n}$ in the standard polytrope equation.
$K=(3\pi^2)^{1/3}\hbar c/(4m_N^{4/3})$ is constant with $\hbar$, $c$ and $m_N$ are reduced Planck constant, speed of light and nucleon mass, respectively.
In addition, taking into account the degenerate pressure of electrons and the finite temperature corrections, the pressure is given by \citep{1990ApJ...353..597B}
\begin{eqnarray}
    P=K(\rho Y_e)^{4/3}\left[1+\frac{2}{3}\left(\frac{s_e}{\pi Y_e}\right)^2\right] \label{eq:LE4}~,
\end{eqnarray}
where $P$, $\rho$, $Y_e$, and $s_e$ are pressure, density, electron fraction, and electronic entropy, respectively.
Assuming that $Y_e$ decreases in correlation with density due to electron capture as $Y_e \propto\rho^\alpha$ and $Y_{e,c}=0.5$
for $\rho_c=10^7{\rm\,g\,cm^{-3}}$ and $Y_{e,c} = 0.42$ for $\rho_c = 10^{10}{\rm \,g\,cm^{-3}}$, we get $P_c \propto\rho^{4(1+\alpha)/3}\sim\rho_c^{1+1/3.3}$, thus $n\sim3.3$ \citep{2018MNRAS.481.3305S}.
Combining Equations \ref{eq:LE2} and \ref{eq:LE4}, we obtained the modified Chandrasekhar mass as
\begin{eqnarray}
    M_{\rm ch} &= 1.09M_\odot \left(\frac{Y_{e,c}}{0.42}\right)^{2}
    \left[1+\frac{2}{3}\left(\frac{s_{e,c}}{\pi Y_e,c}\right)^2\right]^{3/2} 
    \label{eq:Mch1}~,
\end{eqnarray}
where we use $\varphi_n=17.27$ (when $n = 3.3$). 
Equation \ref{eq:Mch1} gives the critical mass above which the central density/pressure cannot support self-gravity anymore.
Finally, we add two corrections from \cite{1990ApJ...353..597B} terms and get,
\begin{eqnarray}
    M_{\rm ch} &= 1.09M_\odot \left(\frac{Y_{e,c}}{0.42}\right)^{2}
    \left[1+\frac{2}{3}\left(\frac{s_{e,c}}{\pi Y_e,c}\right)^2- 0.057 + \frac{s_{e,c}}{A}\right]^{3/2}
    \label{eq:Mch2}~,
\end{eqnarray}
where the third term is due to Coulomb lattice correction and the fourth term is due to ideal ionic gas pressure.

\bibliography{ref}
\bibliographystyle{apj}

\end{document}